\documentclass[3p,twocolumn,sort&compress]{elsarticle}
\usepackage{color}
\usepackage{enumitem}
\usepackage{graphicx}
\usepackage[normalem]{ulem}
\usepackage{comment}
\usepackage{amsmath}
\usepackage{amsfonts}
\usepackage{cuted}
\usepackage{bm}

\makeatletter
\def\ps@pprintTitle{%
 \let\@oddhead\@empty
 \let\@evenhead\@empty
 \let\@oddfoot\@empty
 \let\@evenfoot\@oddfoot
}
\makeatother

\newif\ifHighlitedChanges
\def\ifHighlitedChanges{\iftrue}
\ifHighlitedChanges
  
  \def\STRIKE#1{{\color{red}\sout{#1}}}
\else
  
  \def\STRIKE#1{\relax}
\fi

\begin{document}

\title{Machine learning method to determine concentrations of structural defects in irradiated materials}

\author[add1]{Landon Johnson}
\author[add1,add3]{Walter Malone}
\author[add2]{Jason Rizk}
\author[add1]{Renai Chen}
\author[add1]{Tammie Gibson}
\author[add2]{Michael W. D. Cooper}
\author[add1]{Galen T. Craven}

\address[add1]{Theoretical Division, Los Alamos National Laboratory, Los Alamos, New Mexico 87545}
\address[add2]{Materials Science and Technology Division, Los Alamos National Laboratory, Los Alamos, New Mexico 87545}
\address[add3]{Department of Physics, Tuskegee University, Tuskegee, Alabama 36088}

\begin{abstract}
The formation and subsequent growth of structural defects in an irradiated material can strongly influence 
the material's 
performance in technological and industrial applications. 
Predicting how the growth of defects 
affects material performance is therefore 
a pressing problem in materials science. One common computational approach that is used to examine defect growth 
is cluster dynamics, a method which employs a system of mean-field rate equations to track the time evolution of concentrations of individual defect types. However, the computational complexity of performing cluster dynamics can limit its 
practical 
implementation, specifically in the context of exploring a broad set of physical conditions corresponding to, for example, different temperatures and pressures.
Here, we present a
machine learning
approach to circumvent  the computational challenges of performing cluster dynamics while maintaining high accuracy 
in the prediction of defect concentrations.
Our method is illustrated on the nuclear material uranium nitride but is broadly applicable to other materials.
The developed data-driven method is shown to accurately capture
complex correlations between material properties, temperature, irradiation conditions, and the concentration of defects.
\end{abstract}

\maketitle

\section{Introduction}
Structural defects alter the performance of materials in irradiated environments \cite{Was2016}. 
Predicting how defects form and then grow under a specific set of physical conditions corresponding to, for example, a specific temperature, pressure, and/or fission rate, 
is a problem that has generated significant interest due to its relevance in broad class of technological and industrial applications \cite{MiekeleyFelix1972,Turnbull1982,Matthews1988,Matzke1989,Matzke1990,Sabioni1998,Chaudri2013,Cooper2015,Matthews2019,Matthews2020,Cooper2021,Watkins2021,Rest2019review,Perriot2019,Derlet2020}.
Defects types such as interstitials, bubbles, voids, anti-sites, among others,
can form due to irradiation and alter performance properties \cite{Wolfer1985,Golubov2001,Ortiz2007,Surh2008,Wirth2015,Stewart2018,Kohnert2018}
including thermal conductivity, electrical conductivity, and mechanical response.
Performance alterations in irradiated and nuclear materials due to aging or irradiation-enhanced defect formation \cite{YUN2021100007} 
can significantly reduce the viability of using a specific material due to safety and failure concerns that push an engineered system outside of the safe operating envelope \cite{Stefanescu2023}. 
Therefore, developing theoretical and computational tools to predict how defects form, the rate at which those defects grow, and
how those defects affect the pertinent properties of a material under specific physical conditions are important questions with direct ramifications for technology advancement. Predicting defect growth is particularly important in multiscale modeling applications that use atomistic and/or microscale data to inform macroscale and engineering-level models.

Predicting defect growth rates using computational methods is often a process that involves the implementation of complex computational procedures. For example, defect growth is often predicted by solving large systems of nonlinear differential equations \cite{barbu2007cluster,Kohnert2018} or performing kinetic Monte Carlo simulations \cite{Dai2005LKMC,Dai2006LKMC,Domain2020object}.
Previous theoretical work on defect evolution of irradiated materials has generally focused on the development of systems of master equations that describe how clusters of point defects evolve in time, i.e., cluster dynamics simulations,  \cite{Golubov2001,Wolfer1985,Wirth2015,Bonilla2006,Wirth2015,Gai2015,Li2019}
or on the application of lattice kinetic Monte Carlo methods \cite{Dai2005LKMC,Dai2006LKMC,barbu2007cluster,Domain2020object}.
The specific defect type, or combination of defect types, 
that give rise to structural changes generated by irradiation is typically specific to the material.
In general, the propensity of each defect type to cluster with itself and with other defect types is material dependent \cite{Schaldach2004,Bonilla2006,Thiebaut2007,Sharafat2009,Xu2009,Jeffries2011,Jeffries2018, Xu2013,Tschopp2014,Gai2015,Wang2015,Matthews2019,Matthews2020}.
This poses a problem from a modeling perspective because there is often limited transferability of a developed defect growth model from one material to another. 

One of the primary methods that is used to predict defect evolution in materials is cluster dynamics \cite{Wolfer1985,Golubov2001,Ortiz2007,barbu2007cluster,Surh2008,Wirth2015,Stewart2018,Kohnert2018,Ke2018,Ke2022,Hu2020}.
Cluster dynamics models can be used to predict the outcome of complex physical process that determine the time evolution of defects such as aggregation, recombination, and fragmentation.
In a cluster dynamics simulation, a mean-field method is used to track the time evolution of concentrations or densities of point defects and defect clusters \cite{Kohnert2018,Matthews2019,Matthews2020}.
The mean-field nature of the method arises because the trajectories of each individual defect are not followed, only the evolution of the defect concentrations are followed.
While implementing cluster dynamics is less computationally taxing than 
other approaches, for example kinetic Monte Carlo methods, it can still be computationally prohibitive to perform simulations over a broad set of physical conditions which may be needed to accurately inform macroscopic models in multiscale and multiphysics frameworks, for example, when a cluster dynamics code is used to compute a transport property that is then fed into a macroscale model. 
The speed of performing a multiscale simulation is therefore commonly limited by the model evaluation speed of the atomistic models at lower length scales in the hierarchically multiscale code structure.  
This problem is further pronounced in nonequilibrium atomistic environments, such as a material under irradiation, when the assumptions used to derive equilibrium kinetic theories breakdown \cite{Hu2020,craven15c,craven16c,matyushov16c,craven17a,craven17d}. 

One method that can be applied to microscale material models to increase evaluation speed and improve accuracy is to use data-driven approaches, such as machine learning (ML), to predict defect concentrations and reaction rates.
There have been broad applications of ML methods \cite{jordan2015machine,Mohri2018book,MLbook2} in physicals sciences and engineering \cite{carleo2019machine,zhong2021machine,tarca2007machine,wang2019machine,welborn2018transferability,Kulichenko2021review,butler2018machine,Carrasquilla2017,Carleo2017,Biamonte2017quantum,Deng2017,Liu2019,craven20b,craven20c,boehnlein2022colloquium} including in nuclear materials \cite{Morgan2022review,neudecker2020enhancing,Cai2022MLnuclearfuels,Kautz2019,xu2023advanced,craven2023a}. 
Some of the potential advantages of using data-driven methods are improved accuracy, fast model evaluation, and the ability to capture complex correlations in datasets. Some common disadvantages are that data-driven models typically have limited interpretability, limited transferability, and often generate unphysical predictions when extrapolation outside of the training data is performed.

\begin{figure}[t]
\centering
\includegraphics[width=7.8cm]{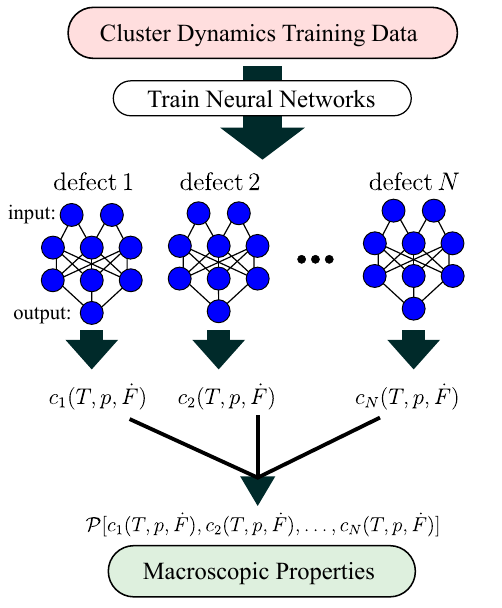}
\caption{\label{fig:TOC}
Schematic diagram of the data-driven methodology developed in this work. 
The defect concentrations obtained from cluster dynamics simulations are used to train a collection of neural networks, one for each of the $N$ defect types tracked in the model. After training, each network can be used to predict the corresponding defect concentration $c$ over a broad range of state points corresponding to different physical conditions. The collection of predicted concentrations can then be used to predict macroscopic properties $\mathcal{P}$ of the material such as atomistic diffusion values and volumetric swelling rates.}
\end{figure}

In this work, we utilize machine learning methods to determine concentrations of point defects and defect clusters in an irradiated material.
Specifically, we present a data-driven approach that uses neural network models trained on limited cluster dynamics simulation data to predict the steady-state concentrations of defects in irradiated materials under a broad set of physical conditions. We illustrate the method on uranium nitride, a potential nuclear fuel.
Our method consists of training a collection of deep neural networks, one network for each defect type in the material model, on data generated using cluster dynamics over a multidimensional space of state configurations. 
The concentrations can then be used to predict macroscopic quantities such as diffusion, volumetric swelling, and others.
An overview of the developed computational workflow is shown in Fig.~\ref{fig:TOC}.
Here, we focus only on point defects and defect clusters, ignoring extended defects. However, the methodology can be modified to include extended defects such as dislocations, voids, and grain boundaries in future work. The primary step in this procedure is to include extended defects in the cluster dynamics model used to generate the training data for the neural networks.
We show that machine learning methodologies can be used to circumvent some of the computational challenges of cluster dynamics that pose problems when building multiscale models that must maintain high accuracy and fast evaluation speed across multiple length scales. 
A comparison between the results of cluster dynamics simulations and the machine learning predictions is presented,
demonstrating that, after training, machine learning can be used to achieve comparable accuracy to cluster dynamics simulations while significantly reducing the computational overhead of determining defect concentrations over various state conditions such as different temperatures, pressures, and fission rates. We find that, after training, the ML methodology can reduce the computational cost needed to sample a new state condition by a factor of $\approx10^4$ in comparison to performing a cluster dynamics simulation. 
The developed methodology allows structural defect formation and growth to be understood 
over a large range of physically-relevant state conditions.

The remainder of this article is organized as follows:
Section~\ref{sec:methods} contains the details of the machine learning methods 
used to examine defect concentrations in irradiated materials.
An overview of how we generated the data used to train and test those methods is also given,
focusing on using calibrated cluster dynamics models to calculate defect concentrations under various 
conditions.
In Section~\ref{sec:results}, the results of the machine learning models are presented.
Concluding remarks are presented in Sec.~\ref{sec:conc}.

\section{Machine Learning Methods \label{sec:methods}}

The data we use to train the collection of neural networks is obtained from cluster dynamics simulations of uranium nitride (UN)
using the model developed in Ref.~\citenum{craven2023b}.
It is important to note that while we illustrate the developed ML method on UN, it is broadly applicable to other irradiated materials
and nuclear fuels.
The developed computational workflow, shown in Fig.~\ref{fig:TOC}, consists of training a collection of neural networks, one for each defect type tracked in the model, to produce the defect concentrations over a broad range of state points corresponding to different temperature $T$, partial pressure of nitrogen $p_{\text{N}_2}$, and fission rate $\dot{F}$. After training, the collection of neural networks can be applied to predict macroscopic properties of the material such as the diffusivity of various species.
The parameters in the UN cluster dynamics model are calibrated to reproduce experimental data. 

Cluster dynamics has been broadly applied to generate defect concentration data for irradiated materials and nuclear fuels \cite{Matthews2019,Matthews2020,Perriot2019,craven2023a,craven2023b,craven2023report}. 
Implementing the cluster dynamics method for a specific material consists of (1) identifying a set of defects in the material
that alter the physical processes under investigation, (2) parameterizing the model, typically using a combination of electronic structure calculations and experimental data, and then (3) solving a typically large system of nonlinear coupled ordinary differential equations, where each equation in the system describes the time evolution of the concentration a 
specific defect type. This procedure can be time-consuming in the model development stage and computationally taxing when used to understand the defect growth under a broad range of physical conditions. In previous work, we utilized ML to accelerate the calibration of cluster dynamics models \cite{craven2023a}. In this work we use ML to the accelerate the computational speed of performing simulations and generating data.

\subsection{Cluster Dynamics \label{sec:cluster}}
The free energy cluster dynamics code {\tt Centipede} \cite{Matthews2019, Matthews2020} 
was applied to incorporate physical parameters into a cluster dynamics model for uranium nitride and then to solve the defect evolution equations. 
Details about the {\tt Centipede} code and the physics behind it can be found in Ref.~\citenum{Matthews2019}.
In a {\tt Centipede} simulation, the concentration $c_d$ of $N$ defect types $d \in \{1,2,\ldots,N\}$ are tracked in time through a system of nonlinear coupled differential equations of the form:
\begin{equation}
\begin{aligned}
\label{eq:CD}
 \frac{dc_1}{dt}  & = \beta_1(t) +\sum_{d} R_{1,d} (c_1,c_{d},D_1, D_{d},T,G) \\
 &\quad-  \sum_{s}S_{1,s} (c_1,c_{s},D_1, T,G),  \\
  \frac{dc_2}{dt}  & = \beta_2(t) +\sum_{d} R_{2,d} (c_2,c_{d},D_2, D_{d},T,G) \\
 &\quad-  \sum_{s}S_{2,s} (c_2,c_{s},D_2, T,G),  \\
 &\ldots& \\
   \frac{dc_N}{dt}  & = \beta_N(t) +\sum_{d} R_{N,d} (c_N,c_{d},D_N, D_{d},T,G) \\
 &\quad-  \sum_{s}S_{N,s} (c_N,c_{s},D_N, T,G),  \\
\end{aligned} 
\end{equation}
where $\beta_d(t)$ is the generation rate of defect $d$ due to irradiation,
$R_{d,d'}$ is the reaction rate between defect types $d$ and $d'$, 
and $S_{d,s}$ is the sink rate between defect type $d$ and sink type $s$.
The sums in Eq.~(\ref{eq:CD}) are self-inclusive and taken over all defect types and all sink types.
The reaction and sink rates depend on the free energy of the system $G$ and temperature of the system $T$.
The reaction rate between defect types $d$ and $d'$ depends on the free energy $G$ and temperature $T$ and also on the concentrations of each defect and the diffusion coefficients $D_d$ and $D_{d'}$ of those defects.
The dependence of the reaction rates on the diffusion coefficients arises because the rate at which the defects move through 
the material lattice dictates the rate at which two defects will come into spatial proximity and interact, potentially combining to form a larger defect type. 
The initial conditions for each simulation are obtained by setting the concentration of each defect type to its equilibrium value. We further assume that the concentrations have no spatial dependence in the material, an assumption that is typical in the performance of cluster dynamics.

In this work, we use cluster dynamics to generate data that is used to train a collection of neural network models, one network for each defect type that is tracked in the simulation.
We specifically use neural networks to calculate the defect concentration values when the system reaches steady-state
under constant source and sink values.
We define the steady-state condition for the material to be reached when 
the rate of change of all the defect concentrations is zero ($\frac{dc_d}{dt} = 0\;\text{for all}\; d$) up to some predefined numerical precision. 
Specifically, we consider a solution converged when $\frac{dc_d}{dt} \leq \mathcal{R}\;\text{for all}\; d$ where $|\mathcal{R}| = 10^{-8}$.
References~\citenum{Matthews2019,Matthews2020} and \citenum{craven2023b} contain further details of the physics models and numerical procedures we use to implement cluster dynamics.

 The free energies and diffusion coefficients for the defects in UN were calculated in previous work \cite{craven2023a} using a combination of density functional theory and empirical potential atomic scale simulations. The point defects considered are vacancies, self-interstitials, anti-sites, and Xe interstitials. Using Kr\"{o}ger-Vink notation~\cite{KROGER1956307}, the point defects listed above are given by V$_\text{U}$, V$_\text{N}$, U$_\text{i}$, N$_\text{i}$, U$_\text{N}$, N$_\text{U}$, and Xe$_\text{i}$. Clusters of vacancies with other vacancies, anti-sites, and Xe defects are also considered, and Table \ref{tab:table1} gives the complete list of defects. The bracket notation ``$\{ \}$" indicates that the defect is a cluster.
 The anti-site defects ($\text{N}_\text{U}$ and $\text{U}_\text{N}$) and the substitutional Xe defects (\{Xe:V$_\text{N}$\} and \{Xe:V$_\text{U}$\}) are assumed to be immobile; the rest of the defects are mobile. The defect mobility is calculated using the atomic scale parameters for migration barriers and attempt frequencies given in Ref.~\citenum{craven2023b}.
 We use a constant sink strength derived from the size and number densities of intragranular bubbles based on microscopy measurements in UN in Ref.~\citenum{Ronchi1978}

  \begin{table*}
  \begin{center}
\caption{\label{tab:table1}List of defect types in the UN model. The Kr\"{o}ger-Vink notation is used for the defect types \cite{KROGER1956307}.}
\begin{tabular}{|c|c|c|}
\hline
defect number& defect type & description\\
\hline
    1           & N$_\text{U}$ & N anti-site \\
\hline    
    2           & \{N$_\text{U}$:V$_\text{U}$\} & N anti-site and U vacancy \\
\hline
    3           & N$_\text{i}$& N interstitial  \\
\hline
    4         &  U$_\text{i}$& U interstitial  \\
\hline
    5    &  U$_\text{N}$  & U anti-site    \\
\hline
    6          &  \{U$_\text{N}$:V$_\text{N}$\}& U anti-site and N vacancy         \\
\hline
    7 &  Xe$_\text{i}$  & Xe interstitial  \\
\hline
    8     &  \{Xe:V$_\text{N}$\} &  Xe and N vacancy\\
\hline
    9         &  \{Xe:2V$_\text{N}$\}&  Xe and double N vacancy   \\
\hline
    10          &  \{Xe:V$_\text{U}$\}& Xe and U vacancy  \\
\hline
    11          &  \{Xe:V$_\text{U}$:V$_\text{N}$\}& Xe and U vacancy and N vacancy    \\
\hline
    12        &  \{Xe:2V$_\text{U}$\}& Xe and double U vacancy  \\
\hline
    13   & V$_\text{N}$& N vacancy      \\
\hline
    14          &  \{2V$_\text{N}$\}&  double N vacancy      \\
\hline
    15 &  V$_\text{U}$&  U vacancy  \\
\hline
    16     & \{V$_\text{U}$:V$_\text{N}$\}& U vacancy and N vacancy   \\
\hline
    17     &  \{2V$_\text{U}$\}& double U vacancy   \\
\hline
\end{tabular}
\end{center}
\end{table*}

\subsection{Training Data \label{sec:test}}

\begin{figure}[b]
\centering
\includegraphics[width=7.8cm]{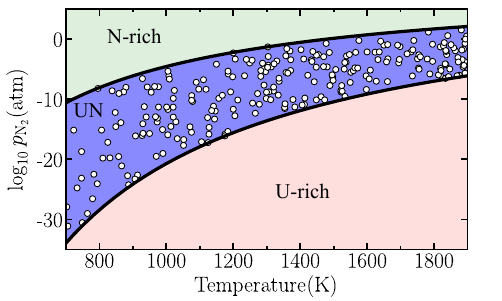}
\caption{\label{fig:phase}
Phase diagram of UN. The white markers are representative state points used in the training data.}
\end{figure}

In order to capture defect behavior under 
a diverse set of physical conditions,
approximately 30,000 {\tt Centipede}  cluster dynamics simulations were performed over varying physical states. 
Each simulation corresponds to a randomly sampled temperature $T$, partial pressure of nitrogen $p_{\text{N}_2}$, and fission rate $\dot{F}$. 
To generate the training data, for each point in the set, the temperature was uniformly sampled between $700\text{K}$ and $2400\text{K}$. After a random temperature was selected, the N$_2$ partial pressure was sampled uniformly on a logarithmic scale
between the temperature-dependent upper and lower bounds defined by $p_\text{upper}(T) = 1.33\times 10^{10} \text{atm}\cdot\exp{\left(\frac{-6.12\text{eV}}{k_\text{B}T}\right)}$ and $p_\text{lower}(T)  = 3.17\times 10^{9}\text{atm}\cdot\exp{\left(\frac{-2.79\text{eV}}{k_\text{B}T}\right)}$. 
These pressure boundaries define the UN-stable region of the phase diagram as shown in Fig.~\ref{fig:phase}.
After the temperature and pressure were determined, the fission rate was sampled uniformly on a logarithmic scale over the range $10^{17}\, \text{fissions} / \text{m}^3 \, \text{s}$ and $10^{21} \, \text{fissions} / \text{m}^3 \, \text{s}$. 
We assume a production of 10,000 Frenkel pairs per fission.
The fisson rate and pressure sampling were performed on a logarithmic scale because the ranges used span several orders of magnitude 
and sampling from the logarithmic scale prevents an under-representation of data at smaller values in the sampling range.
An illustration of the calculated UN phase diagram \cite{craven2023b} and the N-rich, U-rich, and UN-stable regions is shown in Fig.~\ref{fig:phase} with a corresponding scatter plot of a representative set of state conditions in the pressure-temperature plane that were sampled.

{\tt Centipede} calculates the concentration of each defect in terms of site fraction, which is the number of defects in a given volume divided by the number of formula units in a given volume. Therefore throughout this work we express the concentrations of the defects in these relative units.

\subsection{Neural Network Method \label{sec:NN_method}}

\begin{figure}[]
\centering
\includegraphics[width=7.8cm]{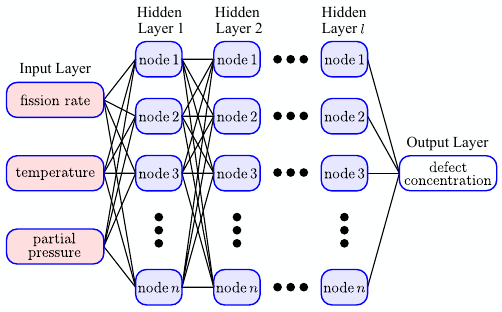}
\caption{\label{fig:NN}
Schematic of the general form of the feed-forward neural network architecture used in this work. The inputs are temperature, partial pressure of N$_2$, and fission rate and the output is the concentration of an individual defect. Each of the $l$ hidden layers consists of $n$ nodes.}
\end{figure}

We constructed our deep neural networks with the \texttt{PyTorch} \cite{Pytorch} machine learning library.
The input layer in the network consists of fission rate, partial pressure of N$_2$, and temperature.  These inputs are successively connected to $l$ hidden layers that are $n$ nodes wide.  As the terminus of the network, the output layer gives an estimation of defect concentration. A different network was trained for each defect type. We denote the specific neural network trained to predict the concentration of defect type $i$ as $\mathcal{N}_i$. A schematic diagram of the neural network architecture used is shown in Fig.~\ref{fig:NN}. All networks are trained using the same architecture.  Each hidden layer $l+1$ node $n$ has the form:
\begin{equation}
x_{l+1,n} = f\bigg(\sum_{n}x_{l,n}W_{l,n}^T+b_{l,n}\bigg).
\end{equation}
In this notation the values of each node $n$ and corresponding layer $x_{l+1}$ are calculated using the values of the subsequent layer $x_{l,n}$, where $W_{l,n}$ and $b_{l,n}$ are the learnable parameters of layer number $l$ and node $n$.  The function $f$ in the above equation is an activation function.  In our case we chose the rectified linear unit function:
\begin{equation}
\text{ReLU}(x) = \max(0,x),
\end{equation}
given that the targets of the neural network, the defect concentrations, are all positive. 
Moreover, the generated dataset was randomly split on an 80:10:10 (training:validation:testing) basis for every network. 
The inputs ($T,p_{N_2},\dot{F}$) and outputs (concentrations) were normalized against the training data according to $(x_i-\bar{x})/\sigma_x$ where $x_i$ is the data point being normalized, $\bar{x}$ is the mean of that data type, and $\sigma_x$ is the standard deviation.
The collection of neural networks $\mathcal{C} = \{\mathcal{N}_1, \mathcal{N}_2,\ldots,\mathcal{N}_N\}$ was used to determine the concentrations
of each defect type using the computational dataflow
\begin{equation}
\underbrace{\{T, p_{\text{N}_2},\dot{F}\}}_\text{inputs} \to \underbrace{\{\mathcal{N}_1, \mathcal{N}_2,\ldots,\mathcal{N}_N\}}_\text{neural networks} \to  \underbrace{\{c_1, c_2,\ldots,c_N\}}_\text{outputs}
\end{equation}
where neural network $\mathcal{N}_1$ generates concentration $c_1$, neural network $\mathcal{N}_2$ generates concentration $c_2$, and so forth.

We performed hyperparameter optimization of the neural network architectures.  
The tested architectures consisted of an input layer (taking values of $T$, $p_{N_2}$, and $\dot{F}$),
$l$ hidden layers, $n$ nodes per hidden layer, and an output layer (the defect concentration), 
with a learning rate $r$.
The hyperparameter optimization was performed using a grid search over combinations of the sets $n\in\{10,15,20,25,30\}$, $l\in\{1,2,3,4\}$, and $r\in\{0.0005,0.001,0.003,0.005\}$. 
After 4,000 epochs, the network architecture with 2 hidden layers of 25 nodes and a learning rate of 0.0005 maintained the lowest error. However, because every network was trained and tested on different random subsets of the {\tt Centipede} data, 
we used the trends in performance to select the best architecture, not the specific error values themselves.
Networks trained with a learning rate of 0.0005 were consistently the best performing. 
A neural network architecture of 2 hidden layers of 25 nodes and a learning rate of 0.0005 was selected for further use after hyperparameter analysis. 
After performing the grid search, the optimal neural network architecture described above was selected and then trained using the {\tt Centipede} cluster dynamics data. 
The same architecture was used for every defect type and each network was trained for 4,000 epochs. The training data inputs and output were log transformed before training each network.

\section{Results \label{sec:results}}

\begin{figure*}[]
\centering
\includegraphics[width=15.6cm]{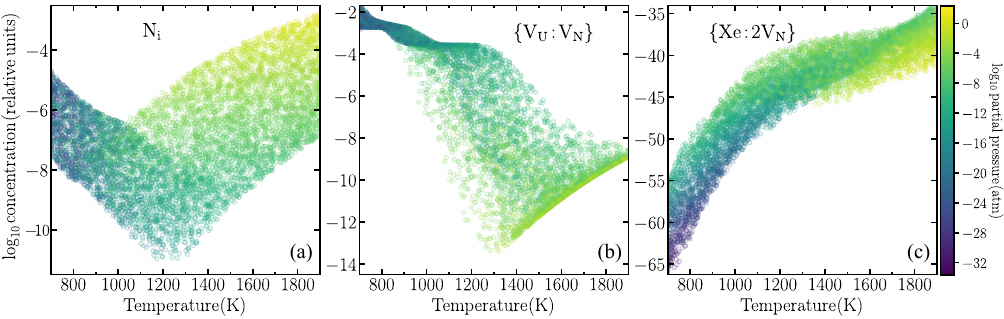}
\caption{\label{fig:CentData}
Concentrations of (a) a  single nitrogen interstitial N$_\text{i}$, (b) a cluster of a U vacancy and a N vacancy \{V$_\text{U}$:V$_\text{N}$\}, and (c) a cluster of a Xe-based double N vacancy \{Xe:2V$_\text{N}$\}. These values were calculated from {\tt Centipede} across varying temperatures, N$_2$ pressures, and fission rates. Temperature is shown on the $x$-axis, partial pressure of N$_2$ is shown with color, blue being the lowest and yellow being the highest pressure. The data shown in each plot is a combination of data from the training, testing, and validation sets.}
\end{figure*}

The concentrations of the $N = 17$ defect types listed in  Table \ref{tab:table1} for UN were calculated with {\tt Centipede} at approximately 30,000 state points. 
The generated concentration training data exhibits strongly nonlinear dependencies on temperature, partial pressure, and fission rate. 
This is illustrated in Fig.~\ref{fig:CentData} 
where it can be seen that the dependence of concentrations on the partial pressure and temperature conditions is highly nonlinear and the overall behavior varies between different defect types. 
Concentration data is shown in Fig.~\ref{fig:CentData} for three defects types: the N interstitial N$_\text{i}$, a cluster of a U vacancy and a N vacancy \{V$_\text{U}$:V$_\text{N}$\}, and a Xe-based double N vacancy \{Xe:2V$_\text{N}$\}. 
These defects were respectively, the best performing, median performing, and worst performing neural networks in terms of percent error as discussed later. 
As shown in Fig.~\ref{fig:CentData}(a), the data for N$_\text{i}$ exhibits two regimes. At temperatures less than $\approx1200 \text{K}$ the distribution of concentrations decreases with increasing temperature. This regime corresponds to the low pressure regime as shown in the color bar of the figure. For temperatures greater than $\approx 1200 \text{K}$, the distribution of concentrations increases nonlinearly as the temperature is increased. This behavior corresponds to higher pressures in the sampling data. 
The data for \{V$_\text{U}$:V$_\text{N}$\} exhibits different behavior as shown in Fig.~\ref{fig:CentData}(b). In the regime of low pressure and low temperature, the concentrations are the highest. As the temperature is increased, the concentration initially begins to decrease but then as higher temperatures are approached the concentrations begin to increase.  
As shown in Fig.~\ref{fig:CentData}(c), the trends are different for \{Xe:2V$_\text{N}$\}. For this defect, no turnover behavior from decreasing to increasing concentration is observed with respect to temperature variation. 
The data shown in Fig.~\ref{fig:CentData} illustrate that the concentration dependence of each defect is highly nonlinear with respect to variation of temperature and partial pressure, and that each defect will, in general, have particular dependencies that are not universal across defect types.

\begin{figure*}[]
\centering
\includegraphics[width=15.6cm]{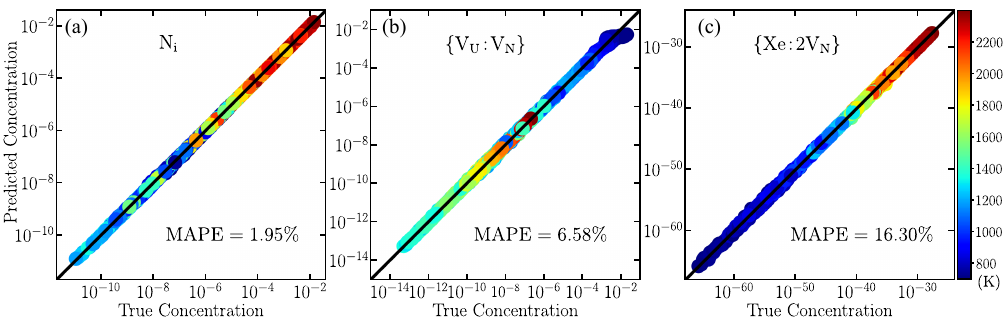}
\caption{\label{fig:parity}
True vs. predicted plots of (a) the best performing neural network corresponding to an N interstital, (b) the median performing neural network corresponding to a cluster of a single U and N vacancy, and (c) the worst performing neural network corresponding to a Xe-based double N vacancy. The color of each marker represents the temperature with scale shown in the color bar on the right. The data shown in each plot is from the testing set.}
\end{figure*}

The performance of the neural network models for some individual defects can be seen in the parity (true vs. predicted) plots in Fig.~\ref{fig:parity}. We again show results for the N interstitial N$_\text{i}$, a cluster of a U vacancy and a N vacancy \{V$_\text{U}$:V$_\text{N}$\}, and a Xe-based double N vacancy \{Xe:2V$_\text{N}$\}. 
The network trained to predict the concentration of a single nitrogen interstitial is the best performing network and delivers the best performance in terms of the lowest mean absolute percent error (MAPE). As shown in  Fig. \ref{fig:parity}(a), the range of concentrations in the testing data for N$_\text{i}$ spans $\approx 10$ orders of magnitude. The MAPE for the cluster consisting of a uranium vacancy and a nitrogen vacancy had the median MAPE, The network trained on a xenon-based double nitrogen vacancies cluster had the largest MAPE of 16.30\%. 
The results for that defect, \{Xe:2V$_\text{N}$\}, are shown in  Fig. \ref{fig:parity}(c). The networks trained on Xe-based defects are generally the east accurate. This is likely due to the fact that the Xe-based defect concentrations span $\approx$30-40 orders of magnitude as opposed to the $\approx$10-20 orders of magnitude for other defect types. This implies that the outputs in the data, i.e., the concentrations, will be significantly more sparse as opposed to other types of defects that do not contain Xe.  

\begin{figure*}[]
\centering
\includegraphics[width=15.6cm]{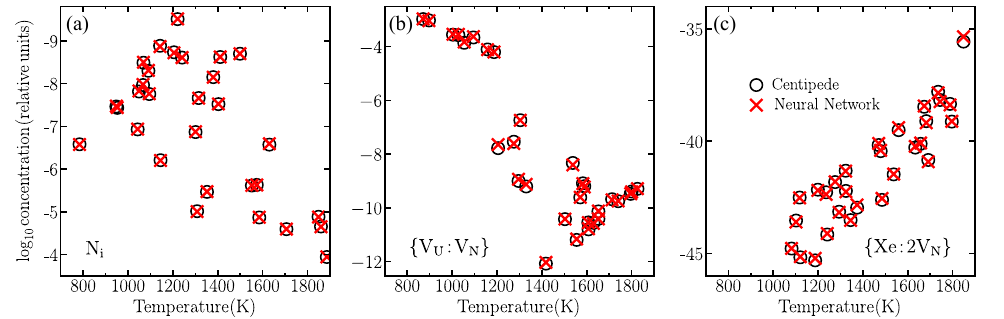}
\caption{\label{fig:comparison}
Comparison of {\tt Centipede} vs. neural network computed defect concentrations for (a) N interstitial, (b) a cluster of a U and N vacancy, and (c) a Xe-based double N vacancy as a function of temperature. These are the same defects/neural networks that are shown in Fig. \ref{fig:parity}.
Each subplot contains results for 30 data points from the test set.
The fission rate in the shown data is constrained to $10^{18.75}-10^{19.25}$ fissions/m$^3$/s
and the nitrogen partial pressure is constrained to the range $10^{-35}-10^{0}$ atm. The data shown in each plot is from the testing set.}
\end{figure*}

Further comparison between the ML model and {\tt Centipede} is shown in Fig.~\ref{fig:comparison}
illustrating that the neural networks are capable of capturing the various non-linear behaviors in the concentrations.
In Fig.~\ref{fig:comparison}, a subset of the total testing data is shown. 
We generate this subset by only including data points 
with fission rate values that fall within the range $10^{18.75}-10^{19.25}$ fissions/m$^3$/s and then randomly selecting 30 data points from the testing set that fall within this range. 
The neural networks are capable of capturing complex nonlinear behaviors.
Specifically, this plot illustrates the ability of the developed methodology to capture nonlinear dependence on temperature. The circular markers are the results from {\tt Centipede} simulations and the red crossed markers are the corresponding predictions generated by the neural network. Excellent agreement is observed between the simulation results and the results generated using ML. It can be seen that the projection of these datapoints with varying pressures and fission rates into the temperature space  shows a nonlinear dependence on temperature. Importantly, we observe no significant outliers in the ML predictions. It can also be seen that there are only small differences between the {\tt Centipede} data and the corresponding ML predictions which implies that the variances in the error distributions generated using the ML method are small.

\begin{figure}[]
\centering
\includegraphics[width=7.8cm]{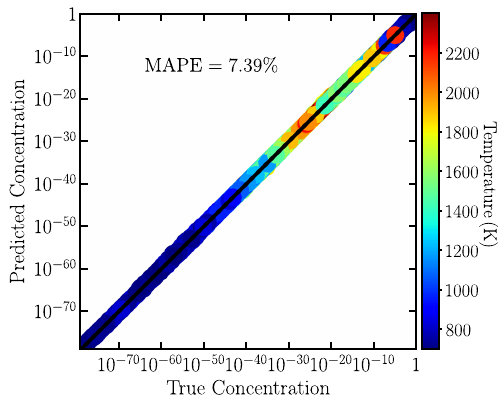}
\caption{\label{fig:allParity}
True concentration vs. predicted concentration combining results from the neural networks for each defect type 
The concentrations are given in values of relative units. The color of each datapoint corresponds to temperature as shown in the color bar on the right. The mean absolute percent error over all defects from this set of networks is 7.39\%. The data shown in each plot is from the testing set.}
\end{figure}

The combined performance across all defect types is shown in the parity plot in Fig.~\ref{fig:allParity}. The predicted values are in strong agreement with the true values, with the ML method developed in this work generating the defect concentrations with a MAPE of 7.39\% when averaged over all defects types. We are therefore able to predict the concentration of defects with low error and with a significant computational savings in comparison to performing cluster dynamics calculations. 
In Fig.~\ref{fig:allParity}, 
the visible temperature trends are mostly representative of Xe-based defect clusters at concentrations lower than approximately $10^{-30}$ and vacancy-type defects at concentrations higher than this. 
This is because the Xe-based defects are the only ones to attain such low concentrations.
Fig.~\ref{fig:allParity} is the primary outcome of this work. 
It illustrates that training a collection of neural networks to predict the concentrations can be used to 
circumvent the computational challenges of determining defect concentrations at steady state in an irradiated material
over a broad set of physical conditions
while capturing 
complex nonlinear correlations between temperature, irradiation conditions, and the concentrations.
Our method is illustrated on the nuclear material uranium nitride but is broadly applicable to other materials.
Timing analysis shows that using the ML-based approach provides a computational increase of a factor $\approx 10^4$ in comparison to performing a new cluster dynamics simulation.
This timing increase is based on a comparison between the START and END time for a
single state point solution using {\tt Centipede}  and the same state point solution using the neural network method. The specific computational increase may be different for different problems. For example, in systems that do not involve a calculation of physical properties like those performed by the {\tt Centipede}  code, a machine learning approach may not provide as significant a reduction in computational cost.

\begin{figure}[t]
\centering
\includegraphics[width=7.8cm]{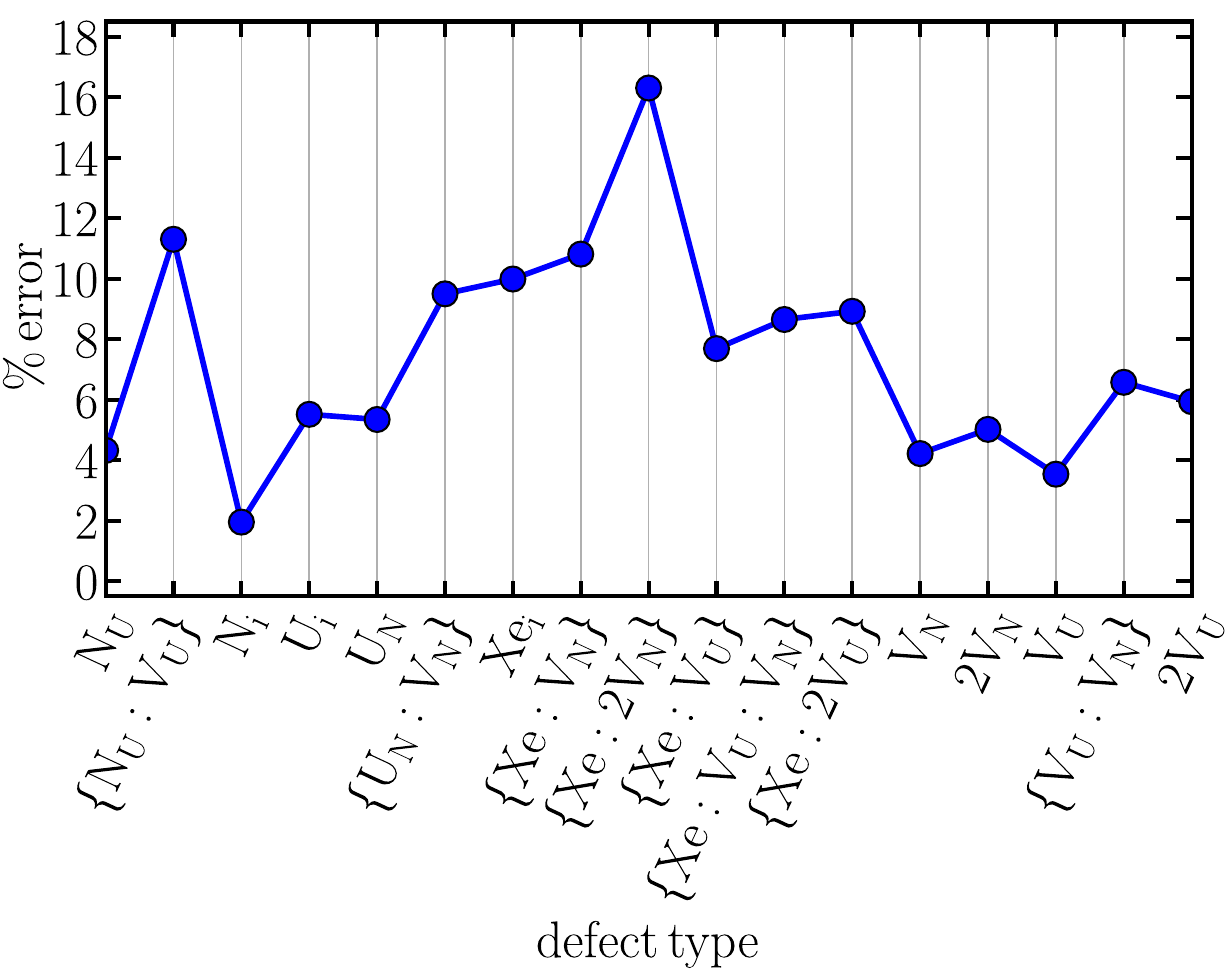}
\caption{\label{fig:defect}
Percent error (MAPE) predicted by the corresponding trained neural network for each defect type in the UN model. The MAPE is calculated over the testing dataset.}
\end{figure}

The MAPE for each defect type is shown in Fig.~\ref{fig:defect}. The MAPE was calculated over the testing data. As discussed previously, the defect types containing Xe are the least accurate networks and generate the largest error. This is in part because the Xe defects have the largest range of distributions in the concentrations over the temperature, pressure, and fission rate regimes that we have sampled and therefore the labeled output data is sparsest for these defect types. 
Another trend that can be observed is that clusters, denoted by the bracket notation ``$\{ \}$", generate higher error than other defects. 
In comparison to the Xe defects, the output concentration range for the other defect types is small and therefore the general trend is that the error is reduced in comparison to clusters that contain Xe. The MAPE averaged over all defects is $\approx 7\%$. 
We have also confirmed that similar trends in the error magnitudes 
can be generated using the developed ML method even when the size of the training dataset is reduced by an order of magnitude.
We found that using a data set of $30,000$ state points generated the targeted predictive accuracy ($<10\%)$ and variance in the neural network models. However, we have examined the performance of the developed approach using smaller data sets and found that fewer state points can be used, particularly if predictions of defect concentrations are needed over a smaller range of physical (temperature, pressure, and fission rate) conditions than what we examine here.

All the results presented in this manuscript constitute interpolation over the training data. The extrapolation performance of the approach will depend on many factors including, for example,  the degree of nonlinearity in the model output and the degree of smoothness in the output. In general, it is expected that the extrapolatory performance of the neural networks trained on  nonlinear CENTIPEDE model outputs will be poor, and we have observed that behavior. However, we have also observed that in some regimes, specifically regimes that depend exponentially on the input variables the model extrapolates well. The cause of this is that these regimes are linear spaces in the log space we work in after log-transforming the data.

 \section{Application to Diffusivity Prediction}

The defect concentrations determined by the ML method can be used to predict other properties of the material that depend on those concentrations, 
for example, the diffusion coefficients of various species in the material \cite{Zhou2021, Zhou2022,craven2023a}.
Here, we calculate the diffusion coefficients of Xe, U, and N in UN to highlight how the developed method can be used to quickly and accurately understand 
important complex physical properties in the material and how the ML method can be applied to supplant the complex and time-consuming task of using human-guided approaches to derive analytical functions to predict material properties.

The diffusion coefficients are calculated using the sum over the product of the relative concentration of each defect contributing to diffusion of a species:
\begin{align}
D_\text{Xe}(T, p_{\text{N}_2}, \dot{F}) &= \frac{\sum\limits_{d \in \mathbf{Xe}} k_\text{B} T g^{(\text{Xe})}_d \, m_d(T) \, c_d(T, p_{\text{N}_2}, \dot{F})}{\sum\limits_{d \in \mathbf{Xe}}c_d(T, p_{\text{N}_2}, \dot{F})}, \\ 
D_\text{N}(T, p_{\text{N}_2}, \dot{F}) &= \sum_{d \in \mathbf{N}} k_\text{B} T g^{(\text{N})}_d \, m_d(T) \, c_d(T, p_{\text{N}_2}, \dot{F}),\\
D_\text{U}(T, p_{\text{N}_2}, \dot{F}) &= \sum_{d \in \mathbf{U}} k_\text{B} T g^{(\text{U})}_d \, m_d(T) \, c_d(T, p_{\text{N}_2}, \dot{F}),  
\end{align}
where $m_d(T)$ is the temperature-dependent mobility of defect $d$ calculated using the methods in Ref.~\citenum{craven2023b}, the sets of defects that are summed over are 
\begin{align}
\mathbf{Xe} &= \nonumber \Big\{\text{Xe}_\text{i}, \{\text{Xe:V}_\text{N}\}, \{\text{Xe:2V}_\text{N}\}, \\
& \qquad \{\text{Xe:V}_\text{U}\},  \{\text{Xe:V}_\text{U}\text{:V}_\text{N}\}, \{\text{Xe:2V}_\text{U}\}\Big\},\\
\mathbf{U} &= \nonumber \Big\{\text{U}_\text{i}, \text{V}_\text{U} , \text{U}_\text{N}, \{\text{N}_\text{U}\text{:V}_\text{U}\}, \\
& \qquad  \{\text{U}_\text{N}\text{:V}_\text{N}\}, \{2\text{V}_\text{U}\}, \{\text{V}_\text{U}\text{:V}_\text{N}\}\Big\},\\
\mathbf{N} &= \nonumber \Big\{\text{N}_\text{i} , \text{V}_\text{N}, \text{N}_\text{U},  \{\text{N}_\text{U}\text{:V}_\text{U}\}, \\
& \qquad \{\text{U}_\text{N}\text{:V}_\text{N}\}, \{2\text{V}_\text{N}\}, \{\text{V}_\text{U}\text{:V}_\text{N}\}\Big\},
\end{align}
and  $g^{(\text{E})}_d:\text{E} \in \{\text{Xe},\text{U}, \text{N}\}$ is the number of contributing defects in defect type $d$ to the diffusion of element $\text{E}$. For all the defect types considered here $g^{(\text{E})}_d = 1$ except for the double vacancy defects where $g^{(\text{N})}_{\{2\text{V}_\text{N}\}} = g^{(\text{U})}_{\{2\text{V}_\text{U}\}} = 2$.
The Xe defect concentrations are normalized using the total Xe concentration.

\begin{figure}[]
\centering
\includegraphics[width=7.8cm]{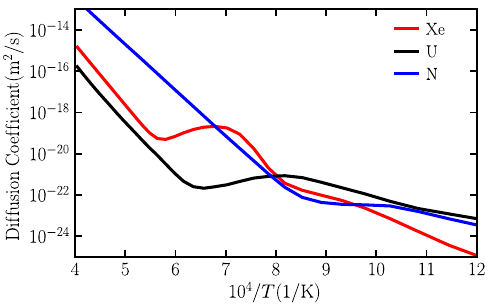}
\caption{\label{fig:all_diff}
Diffusion coefficients predicted by the developed neural network approach for Xe (red), U (black), and N (blue) as a function of inverse temperature. The N$_2$ partial pressure is taken to correspond to the middle of the UN stable region of the phase diagram shown in Fig.~\ref{fig:phase} and the fission rate is $10^{19}\, \text{fissions} / \text{m}^3 \, \text{s}$.}
\end{figure}

Shown in Fig.~\ref{fig:all_diff} are the diffusion coefficients of Xe, U, and N predicted by the developed neural network approach as a function of inverse temperature.
Here, we constrain the partial pressure to be in the middle of the UN phase diagram. This is accomplished by taking a temperature-dependent pressure using the average of the upper $p_\text{upper}$ and lower $p_\text{lower}$ pressure bounds on the UN phase space.
That average pressure is given by $p_\text{mid}(T)  = 6.48\times 10^{9}\text{atm}\cdot\exp{\left(\frac{-4.46\text{eV}}{k_\text{B}T}\right)}$.
The fission rate of $10^{19}\, \text{fissions} / \text{m}^3 \, \text{s}$ is constant over all temperatures and partial pressures.
At higher temperatures, the diffusivity of N dominates 
with the diffusion values for Xe and U being multiple orders of magnitude less than the values for N. 
The predicted diffusion values for each species show the expected Arrenhius-type behavior in the high-temperature limit which
illustrates the ability of the NN approach to capture known physical phenomena. As the temperature is lowered, the Xe defect shows an interesting regime in which the diffusivity increases with decreasing temperature. The U diffusivity also shows this effect but to a lesser degree. 
At the lowest temperature shown in  Fig.~\ref{fig:all_diff}, 
the N and U diffusion coefficients are of a similar magnitude while the Xe value is approximately two orders of magnitude lower than N and U. 
The increased diffusivity for Xe at lower temperatures is caused by an increasing concentration of the highly mobile Xe interstitials instead of immobile substitutional Xe on U sites in that temperature range. For U, this similar behavior is caused by increasing concentration of both U interstitials and U on two vacant N sites instead of the immobile single anti-site.
These results illustrate the use of our NN-based method
to elucidate the behavior of important physical processes, in this case diffusion, in an irradiated material.

\begin{figure}[t]
\centering
\includegraphics[width=7.8cm]{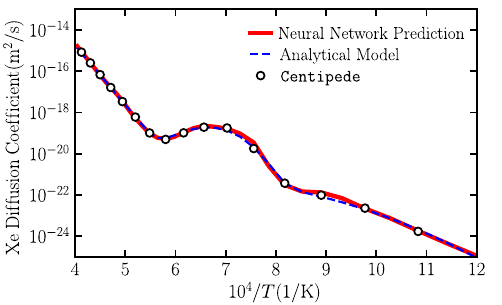}
\caption{\label{fig:Xe_diff}
Xenon diffusion coefficient in UN predicted by the developed neural network method (solid red) and the analytical model in Eq.~(\ref{eq:JasonD}) (dashed blue) as a function of inverse temperature. The results from cluster dynamics calculations performed using the \texttt{Centipede} code are shown as white and black markers. The N$_2$ partial pressure is taken to correspond to the middle of the UN stable region of the phase diagram shown in Fig.~\ref{fig:phase} and the fission rate is $10^{19}\, \text{fissions} / \text{m}^3 \, \text{s}$.}
\end{figure}

Machine learning can provide advantages over human-guided fitting approaches by reducing the complexity of generating an accurate model for a material property
and by increasing the ability to accurately capture complex correlations between material properties.
To illustrate these advantages, we constructed a human-guided function by proposing an analytical function for the Xe diffusion coefficient in UN: 
\begin{align}
\label{eq:JasonD}
D_\text{Xe}(T) &= D_1\exp\left(-\frac{Q_1}{k_\text{B}T}\right) \\ 
\nonumber &+  D_{2} \exp\left(-\frac{Q^{(1)}_{2}}{k_\text{B}T}-\frac{Q^{(2)}_{2}}{(k_\text{B}T)^2}-\frac{Q^{(3)}_{2}}{(k_\text{B}T)^3}\right) \\
\nonumber &+ D_{3} \exp\left(-\frac{Q^{(1)}_{3}}{k_\text{B}T}-\frac{Q^{(2)}_{3}}{(k_\text{B}T)^2}-\frac{Q^{(3)}_{3}}{(k_\text{B}T)^3}\right).
\end{align}
This expression must then be parameterized using a combination of nonlinear regression and by hand tuning of parameters in order to capture the salient physics-informed behavior of Xe diffusion.
The diffusion values generated by {\tt Centipede} and those predicted by the parameterized version of Eq.~(\ref{eq:JasonD}) are shown in Fig.~\ref{fig:Xe_diff}, with excellent agreement observed between the two. 
The major problems with developing these types of empirically-motivated analytical functions are: (1) a proper form for the analytical function must be derived or proposed, which is often a complex procedure, (2) parameterizing that function can require significant time and effort particularly if there are known physics that need to be encoded in the function and its parameters and (3) the function must be reparameterized for each new state point because the parameter values will not be, in general, transferable across different temperatures, pressures, and fission rates. In addition, if an exact value for the particular property is known in a specific limit, the proposed function must also be constrained to agree with that limiting case. These problems can provide significant complexity in the production of analytical forms to fit materials properties.

In comparison, the Xe diffusion values predicted using the ML approach developed here are also in excellent agreement with the {\tt Centipede} cluster dynamics results, as shown in Fig.~\ref{fig:Xe_diff}, 
but the developed method can easily be applied at different state points without reparameterization as that information has been encoded in the weights and biases of the trained collection of neural networks. In addition, because NN are universal function approximators, using the ML approach does not require an appropriate physics-informed functional form to be proposed. This comparison illustrates that ML techniques can efficiently capture nonlinear and intricate correlations among material properties, temperature, and irradiation conditions.

\section{\label{sec:conc}Conclusions}

A machine learning method has been developed to predict the concentrations of point defects and defect clusters in irradiated materials.
Structural defects can significantly alter material properties, and, as such, defect growth poses safety and performance concerns when a material must operate inside a defined operational envelope. 
We have introduced a data-driven approach to predict point defects and defect clusters concentrations under a wide range of physical and irradiation conditions. 
Specifically, we have utilized deep feed-forward neural networks trained on cluster dynamics simulation data to predict defect concentrations. We have illustrated how the concentrations predicted by our model can be applied to determine macroscopic properties like diffusion in a material.
The utility of the method was illustrated on uranium nitride, a potential nuclear fuel, but is broadly applicable to other materials.

The presented results illustrate that machine learning can be used to capture, through a computationally-efficient procedure, 
nonlinear and complex correlations between material properties, temperature, irradiation conditions, and the concentration of defects.
Our method enables the computationally-efficient determination of defect concentrations in irradiated materials.
The presented approach facilitates a more comprehensive understanding of point defect and defect cluster formation and growth across diverse physical conditions.
The machine learning  methodology can be easily modified to include extended defects such as dislocations, voids, and grain boundaries. The primary step in this procedure will be to include extended defects in the cluster dynamics model used to generate the training data. Work in this direction for UN is currently underway.

The developed method is sufficiently generalizable that it can be readily applied to a large set of irradiated materials and nuclear fuels without significant modification. 
The general methodology is transferable across materials. After training a network on data for a specific material, the model is transferable across thermodynamic state points and irradiation conditions for that material provided that the physical state in question is inside the bounds of the training set.
The integration of data-driven approaches with multiscale and multiphysics frameworks could lead to more computationally efficient and accurate simulations workflows that bridge atomistic-scale and engineering-scale models. This will in turn allow for the exploration of a larger design space in technological applications. 
While we have illustrated the method using the input space of temperature, fission rate, and partial pressure, that input space could be extended to include other parameters. Neural network architectures can efficiently handle large input spaces, so we expect that expanding the input space will not limit the applicability of the approach.
We are currently applying techniques such as sensitivity analysis to increase the interpretability of the models.
Overall, this work illustrates the potential of machine learning to address and aid in solving the pressing problem in materials science of predicting structural defects in nuclear and irradiated materials.

\section{Acknowledgments}
This work was supported by the U.S. Department of
Energy through the Los Alamos National Laboratory. Los
Alamos National Laboratory is operated by Triad National
Security, LLC, for the National Nuclear Security Administration of U.S. Department of Energy.
This research was supported by the Laboratory Directed Research and Development program of Los Alamos National Laboratory under project number 20220053DR. 
The computing resources used to perform this
research were provided by the LANL Institutional Computing Program.
Walter Malone acknowledges support from the National Science Foundation: National Science Foundation RISE grant \#2122985. 

\bibliographystyle{apsrev}

\begin{thebibliography}{80}
\expandafter\ifx\csname natexlab\endcsname\relax\def\natexlab#1{#1}\fi
\expandafter\ifx\csname bibnamefont\endcsname\relax
  \def\bibnamefont#1{#1}\fi
\expandafter\ifx\csname bibfnamefont\endcsname\relax
  \def\bibfnamefont#1{#1}\fi
\expandafter\ifx\csname citenamefont\endcsname\relax
  \def\citenamefont#1{#1}\fi
\expandafter\ifx\csname url\endcsname\relax
  \def\url#1{\texttt{#1}}\fi
\expandafter\ifx\csname urlprefix\endcsname\relax\def\urlprefix{URL }\fi
\providecommand{\bibinfo}[2]{#2}
\providecommand{\eprint}[2][]{\url{#2}}

\bibitem[{\citenamefont{Was}(2016)}]{Was2016}
\bibinfo{author}{\bibfnamefont{G.~S.} \bibnamefont{Was}},
  \emph{\bibinfo{title}{Fundamentals of Radiation Materials Science: Metals and
  Alloys}} (\bibinfo{publisher}{Springer}, \bibinfo{year}{2016}).

\bibitem[{\citenamefont{Miekeley and Felix}(1972)}]{MiekeleyFelix1972}
\bibinfo{author}{\bibfnamefont{W.}~\bibnamefont{Miekeley}} \bibnamefont{and}
  \bibinfo{author}{\bibfnamefont{F.}~\bibnamefont{Felix}}, \bibinfo{journal}{J.
  Nucl. Mater.} \textbf{\bibinfo{volume}{42}}, \bibinfo{pages}{297}
  (\bibinfo{year}{1972}), \eprint{doi:10.1016/0022-3115(72)90080-3}.

\bibitem[{\citenamefont{Turnbull et~al.}(1982)\citenamefont{Turnbull, Friskney,
  Findlay, Johnson, and Walter}}]{Turnbull1982}
\bibinfo{author}{\bibfnamefont{J.}~\bibnamefont{Turnbull}},
  \bibinfo{author}{\bibfnamefont{C.}~\bibnamefont{Friskney}},
  \bibinfo{author}{\bibfnamefont{J.}~\bibnamefont{Findlay}},
  \bibinfo{author}{\bibfnamefont{F.}~\bibnamefont{Johnson}}, \bibnamefont{and}
  \bibinfo{author}{\bibfnamefont{A.}~\bibnamefont{Walter}},
  \bibinfo{journal}{J. Nucl. Mater.} \textbf{\bibinfo{volume}{107}},
  \bibinfo{pages}{168} (\bibinfo{year}{1982}),
  \eprint{doi:10.1016/0022-3115(82)90419-6}.

\bibitem[{\citenamefont{Matthews et~al.}(1988)\citenamefont{Matthews,
  Chidester, Hoth, Mason, and Petty}}]{Matthews1988}
\bibinfo{author}{\bibfnamefont{R.}~\bibnamefont{Matthews}},
  \bibinfo{author}{\bibfnamefont{K.}~\bibnamefont{Chidester}},
  \bibinfo{author}{\bibfnamefont{C.}~\bibnamefont{Hoth}},
  \bibinfo{author}{\bibfnamefont{R.}~\bibnamefont{Mason}}, \bibnamefont{and}
  \bibinfo{author}{\bibfnamefont{R.}~\bibnamefont{Petty}}, \bibinfo{journal}{J.
  Nucl. Mater.} \textbf{\bibinfo{volume}{151}}, \bibinfo{pages}{345}
  (\bibinfo{year}{1988}), \eprint{doi:10.1016/0022-3115(88)90029-3}.

\bibitem[{\citenamefont{Matzke}(1989)}]{Matzke1989}
\bibinfo{author}{\bibfnamefont{H.}~\bibnamefont{Matzke}}, in
  \emph{\bibinfo{booktitle}{Studies in Inorganic Chemistry}}, edited by
  \bibinfo{editor}{\bibnamefont{Øivind Johannesen}} \bibnamefont{and}
  \bibinfo{editor}{\bibfnamefont{A.~G.} \bibnamefont{Andersen}}
  (\bibinfo{publisher}{Elsevier}, \bibinfo{year}{1989}),
  vol.~\bibinfo{volume}{9}, pp. \bibinfo{pages}{353--384},
  \eprint{doi:10.1016/B978-0-444-88534-0.50018-7}.

\bibitem[{\citenamefont{Matzke}(1990)}]{Matzke1990}
\bibinfo{author}{\bibfnamefont{H.}~\bibnamefont{Matzke}}, \bibinfo{journal}{J.
  Chem. Soc., Faraday Trans.} \textbf{\bibinfo{volume}{86}},
  \bibinfo{pages}{1243} (\bibinfo{year}{1990}),
  \eprint{doi:10.1039/FT9908601243}.

\bibitem[{\citenamefont{Sabioni et~al.}(1998)\citenamefont{Sabioni, Ferraz, and
  Millot}}]{Sabioni1998}
\bibinfo{author}{\bibfnamefont{A.}~\bibnamefont{Sabioni}},
  \bibinfo{author}{\bibfnamefont{W.}~\bibnamefont{Ferraz}}, \bibnamefont{and}
  \bibinfo{author}{\bibfnamefont{F.}~\bibnamefont{Millot}},
  \bibinfo{journal}{J. Nucl. Mater.} \textbf{\bibinfo{volume}{257}},
  \bibinfo{pages}{180} (\bibinfo{year}{1998}),
  \eprint{doi:10.1016/S0022-3115(98)00482-6}.

\bibitem[{\citenamefont{Chaudri et~al.}(2013)\citenamefont{Chaudri, Tian, Su,
  Zhao, Zhu, Su, and Qiu}}]{Chaudri2013}
\bibinfo{author}{\bibfnamefont{K.~S.} \bibnamefont{Chaudri}},
  \bibinfo{author}{\bibfnamefont{W.}~\bibnamefont{Tian}},
  \bibinfo{author}{\bibfnamefont{Y.}~\bibnamefont{Su}},
  \bibinfo{author}{\bibfnamefont{H.}~\bibnamefont{Zhao}},
  \bibinfo{author}{\bibfnamefont{D.}~\bibnamefont{Zhu}},
  \bibinfo{author}{\bibfnamefont{G.}~\bibnamefont{Su}}, \bibnamefont{and}
  \bibinfo{author}{\bibfnamefont{S.}~\bibnamefont{Qiu}},
  \bibinfo{journal}{Prog. Nucl. Energy} \textbf{\bibinfo{volume}{63}},
  \bibinfo{pages}{57} (\bibinfo{year}{2013}),
  \eprint{doi:10.1016/j.pnucene.2012.11.001}.

\bibitem[{\citenamefont{Cooper et~al.}(2015)\citenamefont{Cooper, Grimes,
  Fitzpatrick, and Chroneos}}]{Cooper2015}
\bibinfo{author}{\bibfnamefont{M.}~\bibnamefont{Cooper}},
  \bibinfo{author}{\bibfnamefont{R.}~\bibnamefont{Grimes}},
  \bibinfo{author}{\bibfnamefont{M.}~\bibnamefont{Fitzpatrick}},
  \bibnamefont{and} \bibinfo{author}{\bibfnamefont{A.}~\bibnamefont{Chroneos}},
  \bibinfo{journal}{Solid State Ion.} \textbf{\bibinfo{volume}{282}},
  \bibinfo{pages}{26} (\bibinfo{year}{2015}),
  \eprint{doi:10.1016/j.ssi.2015.09.006}.

\bibitem[{\citenamefont{Matthews et~al.}(2019)\citenamefont{Matthews, Perriot,
  Cooper, Stanek, and Andersson}}]{Matthews2019}
\bibinfo{author}{\bibfnamefont{C.}~\bibnamefont{Matthews}},
  \bibinfo{author}{\bibfnamefont{R.}~\bibnamefont{Perriot}},
  \bibinfo{author}{\bibfnamefont{M.~W.} \bibnamefont{Cooper}},
  \bibinfo{author}{\bibfnamefont{C.~R.} \bibnamefont{Stanek}},
  \bibnamefont{and} \bibinfo{author}{\bibfnamefont{D.~A.}
  \bibnamefont{Andersson}}, \bibinfo{journal}{J. Nucl. Mater.}
  \textbf{\bibinfo{volume}{527}}, \bibinfo{pages}{151787}
  (\bibinfo{year}{2019}), \eprint{doi:10.1016/j.jnucmat.2019.151787}.

\bibitem[{\citenamefont{Matthews et~al.}(2020)\citenamefont{Matthews, Perriot,
  Cooper, Stanek, and Andersson}}]{Matthews2020}
\bibinfo{author}{\bibfnamefont{C.}~\bibnamefont{Matthews}},
  \bibinfo{author}{\bibfnamefont{R.}~\bibnamefont{Perriot}},
  \bibinfo{author}{\bibfnamefont{M.}~\bibnamefont{Cooper}},
  \bibinfo{author}{\bibfnamefont{C.~R.} \bibnamefont{Stanek}},
  \bibnamefont{and} \bibinfo{author}{\bibfnamefont{D.~A.}
  \bibnamefont{Andersson}}, \bibinfo{journal}{J. Nucl. Mater.}
  \textbf{\bibinfo{volume}{540}}, \bibinfo{pages}{152326}
  (\bibinfo{year}{2020}), \eprint{doi:10.1016/j.jnucmat.2020.152326}.

\bibitem[{\citenamefont{Cooper et~al.}(2021)\citenamefont{Cooper, Pastore, Che,
  Matthews, Forslund, Stanek, Shirvan, Tverberg, Gamble, Mays
  et~al.}}]{Cooper2021}
\bibinfo{author}{\bibfnamefont{M.~W.} \bibnamefont{Cooper}},
  \bibinfo{author}{\bibfnamefont{G.}~\bibnamefont{Pastore}},
  \bibinfo{author}{\bibfnamefont{Y.}~\bibnamefont{Che}},
  \bibinfo{author}{\bibfnamefont{C.}~\bibnamefont{Matthews}},
  \bibinfo{author}{\bibfnamefont{A.}~\bibnamefont{Forslund}},
  \bibinfo{author}{\bibfnamefont{C.~R.} \bibnamefont{Stanek}},
  \bibinfo{author}{\bibfnamefont{K.}~\bibnamefont{Shirvan}},
  \bibinfo{author}{\bibfnamefont{T.}~\bibnamefont{Tverberg}},
  \bibinfo{author}{\bibfnamefont{K.~A.} \bibnamefont{Gamble}},
  \bibinfo{author}{\bibfnamefont{B.}~\bibnamefont{Mays}}, \bibnamefont{et~al.},
  \bibinfo{journal}{Journal of Nuclear Materials}
  \textbf{\bibinfo{volume}{545}}, \bibinfo{pages}{152590}
  (\bibinfo{year}{2021}), \eprint{doi:10.1016/j.jnucmat.2020.152590}.

\bibitem[{\citenamefont{Watkins et~al.}(2021)\citenamefont{Watkins, Gonzales,
  Wagner, Sooby, and Jaques}}]{Watkins2021}
\bibinfo{author}{\bibfnamefont{J.~K.} \bibnamefont{Watkins}},
  \bibinfo{author}{\bibfnamefont{A.}~\bibnamefont{Gonzales}},
  \bibinfo{author}{\bibfnamefont{A.~R.} \bibnamefont{Wagner}},
  \bibinfo{author}{\bibfnamefont{E.~S.} \bibnamefont{Sooby}}, \bibnamefont{and}
  \bibinfo{author}{\bibfnamefont{B.~J.} \bibnamefont{Jaques}},
  \bibinfo{journal}{J. Nucl. Mater.} \textbf{\bibinfo{volume}{553}},
  \bibinfo{pages}{153048} (\bibinfo{year}{2021}),
  \eprint{doi:10.1016/j.jnucmat.2021.153048}.

\bibitem[{\citenamefont{Rest et~al.}(2019)\citenamefont{Rest, Cooper, Spino,
  Turnbull, {Van Uffelen}, and Walker}}]{Rest2019review}
\bibinfo{author}{\bibfnamefont{J.}~\bibnamefont{Rest}},
  \bibinfo{author}{\bibfnamefont{M.}~\bibnamefont{Cooper}},
  \bibinfo{author}{\bibfnamefont{J.}~\bibnamefont{Spino}},
  \bibinfo{author}{\bibfnamefont{J.}~\bibnamefont{Turnbull}},
  \bibinfo{author}{\bibfnamefont{P.}~\bibnamefont{{Van Uffelen}}},
  \bibnamefont{and} \bibinfo{author}{\bibfnamefont{C.}~\bibnamefont{Walker}},
  \bibinfo{journal}{J. Nucl. Mater.} \textbf{\bibinfo{volume}{513}},
  \bibinfo{pages}{310} (\bibinfo{year}{2019}),
  \eprint{doi:10.1016/j.jnucmat.2018.08.019}.

\bibitem[{\citenamefont{Perriot et~al.}(2019)\citenamefont{Perriot, Matthews,
  Cooper, Uberuaga, Stanek, and Andersson}}]{Perriot2019}
\bibinfo{author}{\bibfnamefont{R.}~\bibnamefont{Perriot}},
  \bibinfo{author}{\bibfnamefont{C.}~\bibnamefont{Matthews}},
  \bibinfo{author}{\bibfnamefont{M.~W.} \bibnamefont{Cooper}},
  \bibinfo{author}{\bibfnamefont{B.~P.} \bibnamefont{Uberuaga}},
  \bibinfo{author}{\bibfnamefont{C.~R.} \bibnamefont{Stanek}},
  \bibnamefont{and} \bibinfo{author}{\bibfnamefont{D.~A.}
  \bibnamefont{Andersson}}, \bibinfo{journal}{J. Nucl. Mater.}
  \textbf{\bibinfo{volume}{520}}, \bibinfo{pages}{96} (\bibinfo{year}{2019}),
  \eprint{doi:10.1016/j.jnucmat.2019.03.050}.

\bibitem[{\citenamefont{Derlet and Dudarev}(2020)}]{Derlet2020}
\bibinfo{author}{\bibfnamefont{P.~M.} \bibnamefont{Derlet}} \bibnamefont{and}
  \bibinfo{author}{\bibfnamefont{S.~L.} \bibnamefont{Dudarev}},
  \bibinfo{journal}{Phys. Rev. Mater.} \textbf{\bibinfo{volume}{4}},
  \bibinfo{pages}{023605} (\bibinfo{year}{2020}),
  \eprint{doi:10.1103/PhysRevMaterials.4.023605}.

\bibitem[{\citenamefont{Wehner and Wolfer}(1985)}]{Wolfer1985}
\bibinfo{author}{\bibfnamefont{M.~F.} \bibnamefont{Wehner}} \bibnamefont{and}
  \bibinfo{author}{\bibfnamefont{W.~G.} \bibnamefont{Wolfer}},
  \bibinfo{journal}{Philos. Mag. A} \textbf{\bibinfo{volume}{52}},
  \bibinfo{pages}{189} (\bibinfo{year}{1985}),
  \eprint{doi:10.1080/01418618508237618}.

\bibitem[{\citenamefont{Golubov et~al.}(2001)\citenamefont{Golubov, Ovcharenko,
  Barashev, and Singh}}]{Golubov2001}
\bibinfo{author}{\bibfnamefont{S.~I.} \bibnamefont{Golubov}},
  \bibinfo{author}{\bibfnamefont{A.~M.} \bibnamefont{Ovcharenko}},
  \bibinfo{author}{\bibfnamefont{A.~V.} \bibnamefont{Barashev}},
  \bibnamefont{and} \bibinfo{author}{\bibfnamefont{B.~N.} \bibnamefont{Singh}},
  \bibinfo{journal}{Philos. Mag. A} \textbf{\bibinfo{volume}{81}},
  \bibinfo{pages}{643} (\bibinfo{year}{2001}),
  \eprint{doi:10.1080/01418610108212164}.

\bibitem[{\citenamefont{Ortiz and Caturla}(2007)}]{Ortiz2007}
\bibinfo{author}{\bibfnamefont{C.~J.} \bibnamefont{Ortiz}} \bibnamefont{and}
  \bibinfo{author}{\bibfnamefont{M.~J.} \bibnamefont{Caturla}},
  \bibinfo{journal}{Phys. Rev. B} \textbf{\bibinfo{volume}{75}},
  \bibinfo{pages}{184101} (\bibinfo{year}{2007}),
  \eprint{doi:10.1103/PhysRevB.75.184101}.

\bibitem[{\citenamefont{Surh et~al.}(2008)\citenamefont{Surh, Sturgeon, and
  Wolfer}}]{Surh2008}
\bibinfo{author}{\bibfnamefont{M.~P.} \bibnamefont{Surh}},
  \bibinfo{author}{\bibfnamefont{J.~B.} \bibnamefont{Sturgeon}},
  \bibnamefont{and} \bibinfo{author}{\bibfnamefont{W.~G.}
  \bibnamefont{Wolfer}}, \bibinfo{journal}{J. Nucl. Mater.}
  \textbf{\bibinfo{volume}{378}}, \bibinfo{pages}{86} (\bibinfo{year}{2008}),
  \eprint{doi:10.1016/j.jnucmat.2008.05.009}.

\bibitem[{\citenamefont{Wirth et~al.}(2015)\citenamefont{Wirth, Hu, Kohnert,
  and Xu}}]{Wirth2015}
\bibinfo{author}{\bibfnamefont{B.~D.} \bibnamefont{Wirth}},
  \bibinfo{author}{\bibfnamefont{X.}~\bibnamefont{Hu}},
  \bibinfo{author}{\bibfnamefont{A.}~\bibnamefont{Kohnert}}, \bibnamefont{and}
  \bibinfo{author}{\bibfnamefont{D.}~\bibnamefont{Xu}}, \bibinfo{journal}{J.
  Mater. Res.} \textbf{\bibinfo{volume}{30}}, \bibinfo{pages}{1440–1455}
  (\bibinfo{year}{2015}), \eprint{doi:10.1557/jmr.2015.25}.

\bibitem[{\citenamefont{Stewart et~al.}(2018)\citenamefont{Stewart, Kohnert,
  Capolungo, and Dingreville}}]{Stewart2018}
\bibinfo{author}{\bibfnamefont{J.~A.} \bibnamefont{Stewart}},
  \bibinfo{author}{\bibfnamefont{A.~A.} \bibnamefont{Kohnert}},
  \bibinfo{author}{\bibfnamefont{L.}~\bibnamefont{Capolungo}},
  \bibnamefont{and}
  \bibinfo{author}{\bibfnamefont{R.}~\bibnamefont{Dingreville}},
  \bibinfo{journal}{Compl. Matls. Sci.} \textbf{\bibinfo{volume}{148}},
  \bibinfo{pages}{272 } (\bibinfo{year}{2018}),
  \eprint{doi:10.1016/j.commatsci.2018.02.048}.

\bibitem[{\citenamefont{Kohnert et~al.}(2018)\citenamefont{Kohnert, Wirth, and
  Capolungo}}]{Kohnert2018}
\bibinfo{author}{\bibfnamefont{A.~A.} \bibnamefont{Kohnert}},
  \bibinfo{author}{\bibfnamefont{B.~D.} \bibnamefont{Wirth}}, \bibnamefont{and}
  \bibinfo{author}{\bibfnamefont{L.}~\bibnamefont{Capolungo}},
  \bibinfo{journal}{Compl. Matls. Sci.} \textbf{\bibinfo{volume}{149}},
  \bibinfo{pages}{442} (\bibinfo{year}{2018}),
  \eprint{doi:10.1016/j.commatsci.2018.02.049}.

\bibitem[{\citenamefont{Yun et~al.}(2021)\citenamefont{Yun, Lu, Zhou, Wu, Liu,
  Guo, Shi, and Stubbins}}]{YUN2021100007}
\bibinfo{author}{\bibfnamefont{D.}~\bibnamefont{Yun}},
  \bibinfo{author}{\bibfnamefont{C.}~\bibnamefont{Lu}},
  \bibinfo{author}{\bibfnamefont{Z.}~\bibnamefont{Zhou}},
  \bibinfo{author}{\bibfnamefont{Y.}~\bibnamefont{Wu}},
  \bibinfo{author}{\bibfnamefont{W.}~\bibnamefont{Liu}},
  \bibinfo{author}{\bibfnamefont{S.}~\bibnamefont{Guo}},
  \bibinfo{author}{\bibfnamefont{T.}~\bibnamefont{Shi}}, \bibnamefont{and}
  \bibinfo{author}{\bibfnamefont{J.~F.} \bibnamefont{Stubbins}},
  \bibinfo{journal}{Materials Reports: Energy} \textbf{\bibinfo{volume}{1}},
  \bibinfo{pages}{100007} (\bibinfo{year}{2021}).

\bibitem[{\citenamefont{Stefanescu et~al.}(2023)\citenamefont{Stefanescu,
  Boleininger, and Ma}}]{Stefanescu2023}
\bibinfo{author}{\bibfnamefont{L.}~\bibnamefont{Stefanescu}},
  \bibinfo{author}{\bibfnamefont{M.}~\bibnamefont{Boleininger}},
  \bibnamefont{and} \bibinfo{author}{\bibfnamefont{P.-W.} \bibnamefont{Ma}},
  \bibinfo{journal}{Phys. Rev. Mater.} \textbf{\bibinfo{volume}{7}},
  \bibinfo{pages}{073604} (\bibinfo{year}{2023}),
  \eprint{doi:10.1103/PhysRevMaterials.7.073604}.

\bibitem[{\citenamefont{Barbu and Clouet}(2007)}]{barbu2007cluster}
\bibinfo{author}{\bibfnamefont{A.}~\bibnamefont{Barbu}} \bibnamefont{and}
  \bibinfo{author}{\bibfnamefont{E.}~\bibnamefont{Clouet}}, in
  \emph{\bibinfo{booktitle}{Solid State Phenomena}}
  (\bibinfo{organization}{Trans Tech Publ}, \bibinfo{year}{2007}), vol.
  \bibinfo{volume}{129}, pp. \bibinfo{pages}{51--58}.

\bibitem[{\citenamefont{Dai et~al.}(2005)\citenamefont{Dai, Kanter, Kapur,
  Seider, and Sinno}}]{Dai2005LKMC}
\bibinfo{author}{\bibfnamefont{J.}~\bibnamefont{Dai}},
  \bibinfo{author}{\bibfnamefont{J.~M.} \bibnamefont{Kanter}},
  \bibinfo{author}{\bibfnamefont{S.~S.} \bibnamefont{Kapur}},
  \bibinfo{author}{\bibfnamefont{W.~D.} \bibnamefont{Seider}},
  \bibnamefont{and} \bibinfo{author}{\bibfnamefont{T.}~\bibnamefont{Sinno}},
  \bibinfo{journal}{Phys. Rev. B} \textbf{\bibinfo{volume}{72}},
  \bibinfo{pages}{134102} (\bibinfo{year}{2005}),
  \eprint{doi:10.1103/PhysRevB.72.134102}.

\bibitem[{\citenamefont{Dai et~al.}(2006)\citenamefont{Dai, Seider, and
  Sinno}}]{Dai2006LKMC}
\bibinfo{author}{\bibfnamefont{J.}~\bibnamefont{Dai}},
  \bibinfo{author}{\bibfnamefont{W.~D.} \bibnamefont{Seider}},
  \bibnamefont{and} \bibinfo{author}{\bibfnamefont{T.}~\bibnamefont{Sinno}},
  \bibinfo{journal}{Mol. Simul.} \textbf{\bibinfo{volume}{32}},
  \bibinfo{pages}{305} (\bibinfo{year}{2006}),
  \eprint{doi:10.1080/08927020600586557}.

\bibitem[{\citenamefont{Domain and Becquart}(2020)}]{Domain2020object}
\bibinfo{author}{\bibfnamefont{C.}~\bibnamefont{Domain}} \bibnamefont{and}
  \bibinfo{author}{\bibfnamefont{C.~S.} \bibnamefont{Becquart}},
  \bibinfo{journal}{Handbook of Materials Modeling: Methods: Theory and
  Modeling} pp. \bibinfo{pages}{1287--1312} (\bibinfo{year}{2020}).

\bibitem[{\citenamefont{Bonilla et~al.}(2006)\citenamefont{Bonilla, Carpio,
  Neu, and Wolfer}}]{Bonilla2006}
\bibinfo{author}{\bibfnamefont{L.}~\bibnamefont{Bonilla}},
  \bibinfo{author}{\bibfnamefont{A.}~\bibnamefont{Carpio}},
  \bibinfo{author}{\bibfnamefont{J.}~\bibnamefont{Neu}}, \bibnamefont{and}
  \bibinfo{author}{\bibfnamefont{W.}~\bibnamefont{Wolfer}},
  \bibinfo{journal}{Physica D} \textbf{\bibinfo{volume}{222}},
  \bibinfo{pages}{131} (\bibinfo{year}{2006}),
  \eprint{doi:10.1016/j.physd.2006.07.029}.

\bibitem[{\citenamefont{Gai et~al.}(2015)\citenamefont{Gai, Lazauskas, Smith,
  and Kenny}}]{Gai2015}
\bibinfo{author}{\bibfnamefont{X.}~\bibnamefont{Gai}},
  \bibinfo{author}{\bibfnamefont{T.}~\bibnamefont{Lazauskas}},
  \bibinfo{author}{\bibfnamefont{R.}~\bibnamefont{Smith}}, \bibnamefont{and}
  \bibinfo{author}{\bibfnamefont{S.~D.} \bibnamefont{Kenny}},
  \bibinfo{journal}{J. Nucl. Mater.} \textbf{\bibinfo{volume}{462}},
  \bibinfo{pages}{382} (\bibinfo{year}{2015}),
  \eprint{doi:10.1016/j.jnucmat.2014.10.027}.

\bibitem[{\citenamefont{Li et~al.}(2019)\citenamefont{Li, Li, and
  Han}}]{Li2019}
\bibinfo{author}{\bibfnamefont{S.-H.} \bibnamefont{Li}},
  \bibinfo{author}{\bibfnamefont{J.-T.} \bibnamefont{Li}}, \bibnamefont{and}
  \bibinfo{author}{\bibfnamefont{W.-Z.} \bibnamefont{Han}},
  \bibinfo{journal}{Materials} \textbf{\bibinfo{volume}{12}},
  \bibinfo{pages}{1036} (\bibinfo{year}{2019}),
  \eprint{doi:10.3390/ma12071036}.

\bibitem[{\citenamefont{Schaldach and Wolfer}(2004)}]{Schaldach2004}
\bibinfo{author}{\bibfnamefont{C.}~\bibnamefont{Schaldach}} \bibnamefont{and}
  \bibinfo{author}{\bibfnamefont{W.}~\bibnamefont{Wolfer}}, in
  \emph{\bibinfo{booktitle}{Effects of Radiation on Materials: 21st
  International Symposium}} (\bibinfo{organization}{ASTM International},
  \bibinfo{year}{2004}).

\bibitem[{\citenamefont{Thiebaut et~al.}(2007)\citenamefont{Thiebaut, Baclet,
  Ravat, Giraud, and Julia}}]{Thiebaut2007}
\bibinfo{author}{\bibfnamefont{C.}~\bibnamefont{Thiebaut}},
  \bibinfo{author}{\bibfnamefont{N.}~\bibnamefont{Baclet}},
  \bibinfo{author}{\bibfnamefont{B.}~\bibnamefont{Ravat}},
  \bibinfo{author}{\bibfnamefont{P.}~\bibnamefont{Giraud}}, \bibnamefont{and}
  \bibinfo{author}{\bibfnamefont{P.}~\bibnamefont{Julia}}, \bibinfo{journal}{J.
  Nucl. Mater.} \textbf{\bibinfo{volume}{361}}, \bibinfo{pages}{184}
  (\bibinfo{year}{2007}), \eprint{doi:10.1016/j.jnucmat.2006.12.024}.

\bibitem[{\citenamefont{Sharafat et~al.}(2009)\citenamefont{Sharafat,
  Takahashi, Nagasawa, and Ghoniem}}]{Sharafat2009}
\bibinfo{author}{\bibfnamefont{S.}~\bibnamefont{Sharafat}},
  \bibinfo{author}{\bibfnamefont{A.}~\bibnamefont{Takahashi}},
  \bibinfo{author}{\bibfnamefont{K.}~\bibnamefont{Nagasawa}}, \bibnamefont{and}
  \bibinfo{author}{\bibfnamefont{N.}~\bibnamefont{Ghoniem}},
  \bibinfo{journal}{J. Nucl. Mater.} \textbf{\bibinfo{volume}{389}},
  \bibinfo{pages}{203} (\bibinfo{year}{2009}),
  \eprint{doi:10.1016/j.jnucmat.2009.02.027}.

\bibitem[{\citenamefont{Xu and Wirth}(2009)}]{Xu2009}
\bibinfo{author}{\bibfnamefont{D.}~\bibnamefont{Xu}} \bibnamefont{and}
  \bibinfo{author}{\bibfnamefont{B.~D.} \bibnamefont{Wirth}},
  \bibinfo{journal}{Fusion. Sci. Technol.} \textbf{\bibinfo{volume}{56}},
  \bibinfo{pages}{1064} (\bibinfo{year}{2009}),
  \eprint{doi:10.13182/FST09-A9052}.

\bibitem[{\citenamefont{Jeffries et~al.}(2011)\citenamefont{Jeffries, Wall,
  Moore, and Schwartz}}]{Jeffries2011}
\bibinfo{author}{\bibfnamefont{J.}~\bibnamefont{Jeffries}},
  \bibinfo{author}{\bibfnamefont{M.}~\bibnamefont{Wall}},
  \bibinfo{author}{\bibfnamefont{K.}~\bibnamefont{Moore}}, \bibnamefont{and}
  \bibinfo{author}{\bibfnamefont{A.}~\bibnamefont{Schwartz}},
  \bibinfo{journal}{J. Nucl. Mater.} \textbf{\bibinfo{volume}{410}},
  \bibinfo{pages}{84} (\bibinfo{year}{2011}),
  \eprint{doi:10.1016/j.jnucmat.2011.01.015}.

\bibitem[{\citenamefont{Jeffries et~al.}(2018)\citenamefont{Jeffries, Hammons,
  Willey, Wall, Ruddle, Ilavsky, Allen, and {van Buuren}}}]{Jeffries2018}
\bibinfo{author}{\bibfnamefont{J.}~\bibnamefont{Jeffries}},
  \bibinfo{author}{\bibfnamefont{J.}~\bibnamefont{Hammons}},
  \bibinfo{author}{\bibfnamefont{T.}~\bibnamefont{Willey}},
  \bibinfo{author}{\bibfnamefont{M.}~\bibnamefont{Wall}},
  \bibinfo{author}{\bibfnamefont{D.}~\bibnamefont{Ruddle}},
  \bibinfo{author}{\bibfnamefont{J.}~\bibnamefont{Ilavsky}},
  \bibinfo{author}{\bibfnamefont{P.}~\bibnamefont{Allen}}, \bibnamefont{and}
  \bibinfo{author}{\bibfnamefont{T.}~\bibnamefont{{van Buuren}}},
  \bibinfo{journal}{J. Nucl. Mater.} \textbf{\bibinfo{volume}{498}},
  \bibinfo{pages}{505} (\bibinfo{year}{2018}),
  \eprint{doi:10.1016/j.jnucmat.2017.10.058}.

\bibitem[{\citenamefont{Xu et~al.}(2013)\citenamefont{Xu, Hu, and
  Wirth}}]{Xu2013}
\bibinfo{author}{\bibfnamefont{D.}~\bibnamefont{Xu}},
  \bibinfo{author}{\bibfnamefont{X.}~\bibnamefont{Hu}}, \bibnamefont{and}
  \bibinfo{author}{\bibfnamefont{B.~D.} \bibnamefont{Wirth}},
  \bibinfo{journal}{Appl. Phys. Lett.} \textbf{\bibinfo{volume}{102}},
  \bibinfo{pages}{011904} (\bibinfo{year}{2013}),
  \eprint{doi:10.1063/1.4773876}.

\bibitem[{\citenamefont{Tschopp et~al.}(2014)\citenamefont{Tschopp, Gao, Yang,
  and Solanki}}]{Tschopp2014}
\bibinfo{author}{\bibfnamefont{M.~A.} \bibnamefont{Tschopp}},
  \bibinfo{author}{\bibfnamefont{F.}~\bibnamefont{Gao}},
  \bibinfo{author}{\bibfnamefont{L.}~\bibnamefont{Yang}}, \bibnamefont{and}
  \bibinfo{author}{\bibfnamefont{K.~N.} \bibnamefont{Solanki}},
  \bibinfo{journal}{J. Appl. Phys.} \textbf{\bibinfo{volume}{115}},
  \bibinfo{pages}{033503} (\bibinfo{year}{2014}),
  \eprint{doi:10.1063/1.4861719}.

\bibitem[{\citenamefont{Wang et~al.}(2015)\citenamefont{Wang, Ren, Zhang, Gong,
  Huai, Zhu, Deng, and Hu}}]{Wang2015}
\bibinfo{author}{\bibfnamefont{C.}~\bibnamefont{Wang}},
  \bibinfo{author}{\bibfnamefont{C.}~\bibnamefont{Ren}},
  \bibinfo{author}{\bibfnamefont{W.}~\bibnamefont{Zhang}},
  \bibinfo{author}{\bibfnamefont{H.}~\bibnamefont{Gong}},
  \bibinfo{author}{\bibfnamefont{P.}~\bibnamefont{Huai}},
  \bibinfo{author}{\bibfnamefont{Z.}~\bibnamefont{Zhu}},
  \bibinfo{author}{\bibfnamefont{H.}~\bibnamefont{Deng}}, \bibnamefont{and}
  \bibinfo{author}{\bibfnamefont{W.}~\bibnamefont{Hu}},
  \bibinfo{journal}{Compl. Matls. Sci.} \textbf{\bibinfo{volume}{107}},
  \bibinfo{pages}{54} (\bibinfo{year}{2015}),
  \eprint{doi:10.1016/j.commatsci.2015.05.017}.

\bibitem[{\citenamefont{Ke et~al.}(2018)\citenamefont{Ke, Ke, Odette, and
  Morgan}}]{Ke2018}
\bibinfo{author}{\bibfnamefont{J.-H.} \bibnamefont{Ke}},
  \bibinfo{author}{\bibfnamefont{H.}~\bibnamefont{Ke}},
  \bibinfo{author}{\bibfnamefont{G.~R.} \bibnamefont{Odette}},
  \bibnamefont{and} \bibinfo{author}{\bibfnamefont{D.}~\bibnamefont{Morgan}},
  \bibinfo{journal}{J. Nucl. Mater.} \textbf{\bibinfo{volume}{498}},
  \bibinfo{pages}{83} (\bibinfo{year}{2018}),
  \eprint{doi:10.1016/j.jnucmat.2017.10.008}.

\bibitem[{\citenamefont{Ke and Spencer}(2022)}]{Ke2022}
\bibinfo{author}{\bibfnamefont{J.-H.} \bibnamefont{Ke}} \bibnamefont{and}
  \bibinfo{author}{\bibfnamefont{B.~W.} \bibnamefont{Spencer}},
  \bibinfo{journal}{J. Nucl. Mater.} \textbf{\bibinfo{volume}{569}},
  \bibinfo{pages}{153910} (\bibinfo{year}{2022}),
  \eprint{doi:10.1016/j.jnucmat.2022.153910}.

\bibitem[{\citenamefont{Hu et~al.}(2020)\citenamefont{Hu, Setyawan, Beeler,
  Gan, and Burkes}}]{Hu2020}
\bibinfo{author}{\bibfnamefont{S.}~\bibnamefont{Hu}},
  \bibinfo{author}{\bibfnamefont{W.}~\bibnamefont{Setyawan}},
  \bibinfo{author}{\bibfnamefont{B.~W.} \bibnamefont{Beeler}},
  \bibinfo{author}{\bibfnamefont{J.}~\bibnamefont{Gan}}, \bibnamefont{and}
  \bibinfo{author}{\bibfnamefont{D.~E.} \bibnamefont{Burkes}},
  \bibinfo{journal}{J. Nucl. Mater.} \textbf{\bibinfo{volume}{542}},
  \bibinfo{pages}{152441} (\bibinfo{year}{2020}),
  \eprint{doi:10.1016/j.jnucmat.2020.152441}.

\bibitem[{\citenamefont{Craven and Hernandez}(2015)}]{craven15c}
\bibinfo{author}{\bibfnamefont{G.~T.} \bibnamefont{Craven}} \bibnamefont{and}
  \bibinfo{author}{\bibfnamefont{R.}~\bibnamefont{Hernandez}},
  \bibinfo{journal}{Phys. Rev. Lett.} \textbf{\bibinfo{volume}{115}},
  \bibinfo{pages}{148301} (\bibinfo{year}{2015}),
  \eprint{doi:10.1103/PhysRevLett.115.148301}.

\bibitem[{\citenamefont{Craven and Nitzan}(2016)}]{craven16c}
\bibinfo{author}{\bibfnamefont{G.~T.} \bibnamefont{Craven}} \bibnamefont{and}
  \bibinfo{author}{\bibfnamefont{A.}~\bibnamefont{Nitzan}},
  \bibinfo{journal}{Proc. Natl. Acad. Sci.} \textbf{\bibinfo{volume}{113}},
  \bibinfo{pages}{9421} (\bibinfo{year}{2016}),
  \eprint{doi:10.1073/pnas.1609141113}.

\bibitem[{\citenamefont{Matyushov}(2016)}]{matyushov16c}
\bibinfo{author}{\bibfnamefont{D.~V.} \bibnamefont{Matyushov}},
  \bibinfo{journal}{Proc. Natl. Acad. Sci.} \textbf{\bibinfo{volume}{113}},
  \bibinfo{pages}{9401} (\bibinfo{year}{2016}),
  \eprint{doi:10.1073/pnas.1610542113}.

\bibitem[{\citenamefont{Craven and Nitzan}(2017)}]{craven17a}
\bibinfo{author}{\bibfnamefont{G.~T.} \bibnamefont{Craven}} \bibnamefont{and}
  \bibinfo{author}{\bibfnamefont{A.}~\bibnamefont{Nitzan}},
  \bibinfo{journal}{J. Chem. Phys.} \textbf{\bibinfo{volume}{146}},
  \bibinfo{pages}{092305} (\bibinfo{year}{2017}),
  \eprint{doi:10.1063/1.4971293}.

\bibitem[{\citenamefont{Craven et~al.}(2017)\citenamefont{Craven, Junginger,
  and Hernandez}}]{craven17d}
\bibinfo{author}{\bibfnamefont{G.~T.} \bibnamefont{Craven}},
  \bibinfo{author}{\bibfnamefont{A.}~\bibnamefont{Junginger}},
  \bibnamefont{and}
  \bibinfo{author}{\bibfnamefont{R.}~\bibnamefont{Hernandez}},
  \bibinfo{journal}{Phys. Rev. E} \textbf{\bibinfo{volume}{96}},
  \bibinfo{pages}{022222} (\bibinfo{year}{2017}),
  \eprint{doi:10.1103/PhysRevE.96.022222}.

\bibitem[{\citenamefont{Jordan and Mitchell}(2015)}]{jordan2015machine}
\bibinfo{author}{\bibfnamefont{M.~I.} \bibnamefont{Jordan}} \bibnamefont{and}
  \bibinfo{author}{\bibfnamefont{T.~M.} \bibnamefont{Mitchell}},
  \bibinfo{journal}{Science} \textbf{\bibinfo{volume}{349}},
  \bibinfo{pages}{255} (\bibinfo{year}{2015}),
  \eprint{doi:10.1126/science.aaa8415}.

\bibitem[{\citenamefont{Mohri et~al.}(2018)\citenamefont{Mohri, Rostamizadeh,
  and Talwalkar}}]{Mohri2018book}
\bibinfo{author}{\bibfnamefont{M.}~\bibnamefont{Mohri}},
  \bibinfo{author}{\bibfnamefont{A.}~\bibnamefont{Rostamizadeh}},
  \bibnamefont{and}
  \bibinfo{author}{\bibfnamefont{A.}~\bibnamefont{Talwalkar}},
  \emph{\bibinfo{title}{Foundations of machine learning}}
  (\bibinfo{publisher}{MIT press}, \bibinfo{year}{2018}).

\bibitem[{\citenamefont{Friedman et~al.}(2001)\citenamefont{Friedman, Hastie,
  and Tibshirani}}]{MLbook2}
\bibinfo{author}{\bibfnamefont{J.}~\bibnamefont{Friedman}},
  \bibinfo{author}{\bibfnamefont{T.}~\bibnamefont{Hastie}}, \bibnamefont{and}
  \bibinfo{author}{\bibfnamefont{R.}~\bibnamefont{Tibshirani}},
  \emph{\bibinfo{title}{The Elements of Statistical Learning}}
  (\bibinfo{publisher}{Springer, New York}, \bibinfo{year}{2001}).

\bibitem[{\citenamefont{Carleo et~al.}(2019)\citenamefont{Carleo, Cirac,
  Cranmer, Daudet, Schuld, Tishby, Vogt-Maranto, and
  Zdeborov{\'a}}}]{carleo2019machine}
\bibinfo{author}{\bibfnamefont{G.}~\bibnamefont{Carleo}},
  \bibinfo{author}{\bibfnamefont{I.}~\bibnamefont{Cirac}},
  \bibinfo{author}{\bibfnamefont{K.}~\bibnamefont{Cranmer}},
  \bibinfo{author}{\bibfnamefont{L.}~\bibnamefont{Daudet}},
  \bibinfo{author}{\bibfnamefont{M.}~\bibnamefont{Schuld}},
  \bibinfo{author}{\bibfnamefont{N.}~\bibnamefont{Tishby}},
  \bibinfo{author}{\bibfnamefont{L.}~\bibnamefont{Vogt-Maranto}},
  \bibnamefont{and}
  \bibinfo{author}{\bibfnamefont{L.}~\bibnamefont{Zdeborov{\'a}}},
  \bibinfo{journal}{Rev. Mod. Phys.} \textbf{\bibinfo{volume}{91}},
  \bibinfo{pages}{045002} (\bibinfo{year}{2019}),
  \eprint{doi:10.1103/RevModPhys.91.045002}.

\bibitem[{\citenamefont{Zhong et~al.}(2021)\citenamefont{Zhong, Zhang, Bagheri,
  Burken, Gu, Li, Ma, Marrone, Ren, Schrier et~al.}}]{zhong2021machine}
\bibinfo{author}{\bibfnamefont{S.}~\bibnamefont{Zhong}},
  \bibinfo{author}{\bibfnamefont{K.}~\bibnamefont{Zhang}},
  \bibinfo{author}{\bibfnamefont{M.}~\bibnamefont{Bagheri}},
  \bibinfo{author}{\bibfnamefont{J.~G.} \bibnamefont{Burken}},
  \bibinfo{author}{\bibfnamefont{A.}~\bibnamefont{Gu}},
  \bibinfo{author}{\bibfnamefont{B.}~\bibnamefont{Li}},
  \bibinfo{author}{\bibfnamefont{X.}~\bibnamefont{Ma}},
  \bibinfo{author}{\bibfnamefont{B.~L.} \bibnamefont{Marrone}},
  \bibinfo{author}{\bibfnamefont{Z.~J.} \bibnamefont{Ren}},
  \bibinfo{author}{\bibfnamefont{J.}~\bibnamefont{Schrier}},
  \bibnamefont{et~al.}, \bibinfo{journal}{Environ. Sci. amp; Technol.}
  \textbf{\bibinfo{volume}{55}}, \bibinfo{pages}{12741} (\bibinfo{year}{2021}),
  \eprint{doi:10.1021/acs.est.1c01339}.

\bibitem[{\citenamefont{Tarca et~al.}(2007)\citenamefont{Tarca, Carey, Chen,
  Romero, and Dr{\u{a}}ghici}}]{tarca2007machine}
\bibinfo{author}{\bibfnamefont{A.~L.} \bibnamefont{Tarca}},
  \bibinfo{author}{\bibfnamefont{V.~J.} \bibnamefont{Carey}},
  \bibinfo{author}{\bibfnamefont{X.-w.} \bibnamefont{Chen}},
  \bibinfo{author}{\bibfnamefont{R.}~\bibnamefont{Romero}}, \bibnamefont{and}
  \bibinfo{author}{\bibfnamefont{S.}~\bibnamefont{Dr{\u{a}}ghici}},
  \bibinfo{journal}{PLoS Comp. Biol.} \textbf{\bibinfo{volume}{3}},
  \bibinfo{pages}{e116} (\bibinfo{year}{2007}),
  \eprint{doi:10.1371/journal.pcbi.0030116}.

\bibitem[{\citenamefont{Wang et~al.}(2019)\citenamefont{Wang, Olsson, Wehmeyer,
  P{\'e}rez, Charron, De~Fabritiis, No{\'e}, and Clementi}}]{wang2019machine}
\bibinfo{author}{\bibfnamefont{J.}~\bibnamefont{Wang}},
  \bibinfo{author}{\bibfnamefont{S.}~\bibnamefont{Olsson}},
  \bibinfo{author}{\bibfnamefont{C.}~\bibnamefont{Wehmeyer}},
  \bibinfo{author}{\bibfnamefont{A.}~\bibnamefont{P{\'e}rez}},
  \bibinfo{author}{\bibfnamefont{N.~E.} \bibnamefont{Charron}},
  \bibinfo{author}{\bibfnamefont{G.}~\bibnamefont{De~Fabritiis}},
  \bibinfo{author}{\bibfnamefont{F.}~\bibnamefont{No{\'e}}}, \bibnamefont{and}
  \bibinfo{author}{\bibfnamefont{C.}~\bibnamefont{Clementi}},
  \bibinfo{journal}{ACS Cent. Sci.} \textbf{\bibinfo{volume}{5}},
  \bibinfo{pages}{755} (\bibinfo{year}{2019}),
  \eprint{doi:10.1021/acscentsci.8b00913}.

\bibitem[{\citenamefont{Welborn et~al.}(2018)\citenamefont{Welborn, Cheng, and
  Miller~III}}]{welborn2018transferability}
\bibinfo{author}{\bibfnamefont{M.}~\bibnamefont{Welborn}},
  \bibinfo{author}{\bibfnamefont{L.}~\bibnamefont{Cheng}}, \bibnamefont{and}
  \bibinfo{author}{\bibfnamefont{T.~F.} \bibnamefont{Miller~III}},
  \bibinfo{journal}{J. Chem. Theory Comput.} \textbf{\bibinfo{volume}{14}},
  \bibinfo{pages}{4772} (\bibinfo{year}{2018}),
  \eprint{doi:10.1021/acs.jctc.8b00636}.

\bibitem[{\citenamefont{Kulichenko et~al.}(2021)\citenamefont{Kulichenko,
  Smith, Nebgen, Li, Fedik, Boldyrev, Lubbers, Barros, and
  Tretiak}}]{Kulichenko2021review}
\bibinfo{author}{\bibfnamefont{M.}~\bibnamefont{Kulichenko}},
  \bibinfo{author}{\bibfnamefont{J.~S.} \bibnamefont{Smith}},
  \bibinfo{author}{\bibfnamefont{B.}~\bibnamefont{Nebgen}},
  \bibinfo{author}{\bibfnamefont{Y.~W.} \bibnamefont{Li}},
  \bibinfo{author}{\bibfnamefont{N.}~\bibnamefont{Fedik}},
  \bibinfo{author}{\bibfnamefont{A.~I.} \bibnamefont{Boldyrev}},
  \bibinfo{author}{\bibfnamefont{N.}~\bibnamefont{Lubbers}},
  \bibinfo{author}{\bibfnamefont{K.}~\bibnamefont{Barros}}, \bibnamefont{and}
  \bibinfo{author}{\bibfnamefont{S.}~\bibnamefont{Tretiak}},
  \bibinfo{journal}{J. Phys. Chem. Lett.} \textbf{\bibinfo{volume}{12}},
  \bibinfo{pages}{6227} (\bibinfo{year}{2021}),
  \eprint{doi:10.1021/acs.jpclett.1c01357}.

\bibitem[{\citenamefont{Butler et~al.}(2018)\citenamefont{Butler, Davies,
  Cartwright, Isayev, and Walsh}}]{butler2018machine}
\bibinfo{author}{\bibfnamefont{K.~T.} \bibnamefont{Butler}},
  \bibinfo{author}{\bibfnamefont{D.~W.} \bibnamefont{Davies}},
  \bibinfo{author}{\bibfnamefont{H.}~\bibnamefont{Cartwright}},
  \bibinfo{author}{\bibfnamefont{O.}~\bibnamefont{Isayev}}, \bibnamefont{and}
  \bibinfo{author}{\bibfnamefont{A.}~\bibnamefont{Walsh}},
  \bibinfo{journal}{Nature} \textbf{\bibinfo{volume}{559}},
  \bibinfo{pages}{547} (\bibinfo{year}{2018}),
  \eprint{doi:10.1038/s41586-018-0337-2}.

\bibitem[{\citenamefont{Carrasquilla and Melko}(2017)}]{Carrasquilla2017}
\bibinfo{author}{\bibfnamefont{J.}~\bibnamefont{Carrasquilla}}
  \bibnamefont{and} \bibinfo{author}{\bibfnamefont{R.~G.} \bibnamefont{Melko}},
  \bibinfo{journal}{Nature Phys.} \textbf{\bibinfo{volume}{13}},
  \bibinfo{pages}{431} (\bibinfo{year}{2017}), \eprint{doi:10.1038/nphys4035}.

\bibitem[{\citenamefont{Carleo and Troyer}(2017)}]{Carleo2017}
\bibinfo{author}{\bibfnamefont{G.}~\bibnamefont{Carleo}} \bibnamefont{and}
  \bibinfo{author}{\bibfnamefont{M.}~\bibnamefont{Troyer}},
  \bibinfo{journal}{Science} \textbf{\bibinfo{volume}{355}},
  \bibinfo{pages}{602} (\bibinfo{year}{2017}),
  \eprint{doi:10.1126/science.aag2302}.

\bibitem[{\citenamefont{Biamonte et~al.}(2017)\citenamefont{Biamonte, Wittek,
  Pancotti, Rebentrost, Wiebe, and Lloyd}}]{Biamonte2017quantum}
\bibinfo{author}{\bibfnamefont{J.}~\bibnamefont{Biamonte}},
  \bibinfo{author}{\bibfnamefont{P.}~\bibnamefont{Wittek}},
  \bibinfo{author}{\bibfnamefont{N.}~\bibnamefont{Pancotti}},
  \bibinfo{author}{\bibfnamefont{P.}~\bibnamefont{Rebentrost}},
  \bibinfo{author}{\bibfnamefont{N.}~\bibnamefont{Wiebe}}, \bibnamefont{and}
  \bibinfo{author}{\bibfnamefont{S.}~\bibnamefont{Lloyd}},
  \bibinfo{journal}{Nature} \textbf{\bibinfo{volume}{549}},
  \bibinfo{pages}{195} (\bibinfo{year}{2017}),
  \eprint{doi:10.1038/nature23474}.

\bibitem[{\citenamefont{Deng et~al.}(2017)\citenamefont{Deng, Li, and
  Das~Sarma}}]{Deng2017}
\bibinfo{author}{\bibfnamefont{D.-L.} \bibnamefont{Deng}},
  \bibinfo{author}{\bibfnamefont{X.}~\bibnamefont{Li}}, \bibnamefont{and}
  \bibinfo{author}{\bibfnamefont{S.}~\bibnamefont{Das~Sarma}},
  \bibinfo{journal}{Phys. Rev. X} \textbf{\bibinfo{volume}{7}},
  \bibinfo{pages}{021021} (\bibinfo{year}{2017}),
  \eprint{doi:10.1103/PhysRevX.7.021021}.

\bibitem[{\citenamefont{Liu et~al.}(2019)\citenamefont{Liu, Hong, and
  Cao}}]{Liu2019}
\bibinfo{author}{\bibfnamefont{Y.}~\bibnamefont{Liu}},
  \bibinfo{author}{\bibfnamefont{W.}~\bibnamefont{Hong}}, \bibnamefont{and}
  \bibinfo{author}{\bibfnamefont{B.}~\bibnamefont{Cao}},
  \bibinfo{journal}{Energy} \textbf{\bibinfo{volume}{188}},
  \bibinfo{pages}{116091} (\bibinfo{year}{2019}), ISSN
  \bibinfo{issn}{0360-5442}, \eprint{doi:10.1016/j.energy.2019.116091}.

\bibitem[{\citenamefont{Craven et~al.}(2020{\natexlab{a}})\citenamefont{Craven,
  Lubbers, Barros, and Tretiak}}]{craven20b}
\bibinfo{author}{\bibfnamefont{G.~T.} \bibnamefont{Craven}},
  \bibinfo{author}{\bibfnamefont{N.}~\bibnamefont{Lubbers}},
  \bibinfo{author}{\bibfnamefont{K.}~\bibnamefont{Barros}}, \bibnamefont{and}
  \bibinfo{author}{\bibfnamefont{S.}~\bibnamefont{Tretiak}},
  \bibinfo{journal}{J. Phys. Chem. Lett.} \textbf{\bibinfo{volume}{11}},
  \bibinfo{pages}{4372–4378} (\bibinfo{year}{2020}{\natexlab{a}}),
  \eprint{doi:10.1021/acs.jpclett.0c00627}.

\bibitem[{\citenamefont{Craven et~al.}(2020{\natexlab{b}})\citenamefont{Craven,
  Lubbers, Barros, and Tretiak}}]{craven20c}
\bibinfo{author}{\bibfnamefont{G.~T.} \bibnamefont{Craven}},
  \bibinfo{author}{\bibfnamefont{N.}~\bibnamefont{Lubbers}},
  \bibinfo{author}{\bibfnamefont{K.}~\bibnamefont{Barros}}, \bibnamefont{and}
  \bibinfo{author}{\bibfnamefont{S.}~\bibnamefont{Tretiak}},
  \bibinfo{journal}{J. Chem. Phys.} \textbf{\bibinfo{volume}{153}},
  \bibinfo{pages}{104502} (\bibinfo{year}{2020}{\natexlab{b}}),
  \eprint{doi:10.1063/5.0017894}.

\bibitem[{\citenamefont{Boehnlein et~al.}(2022)\citenamefont{Boehnlein,
  Diefenthaler, Sato, Schram, Ziegler, Fanelli, Hjorth-Jensen, Horn, Kuchera,
  Lee et~al.}}]{boehnlein2022colloquium}
\bibinfo{author}{\bibfnamefont{A.}~\bibnamefont{Boehnlein}},
  \bibinfo{author}{\bibfnamefont{M.}~\bibnamefont{Diefenthaler}},
  \bibinfo{author}{\bibfnamefont{N.}~\bibnamefont{Sato}},
  \bibinfo{author}{\bibfnamefont{M.}~\bibnamefont{Schram}},
  \bibinfo{author}{\bibfnamefont{V.}~\bibnamefont{Ziegler}},
  \bibinfo{author}{\bibfnamefont{C.}~\bibnamefont{Fanelli}},
  \bibinfo{author}{\bibfnamefont{M.}~\bibnamefont{Hjorth-Jensen}},
  \bibinfo{author}{\bibfnamefont{T.}~\bibnamefont{Horn}},
  \bibinfo{author}{\bibfnamefont{M.~P.} \bibnamefont{Kuchera}},
  \bibinfo{author}{\bibfnamefont{D.}~\bibnamefont{Lee}}, \bibnamefont{et~al.},
  \bibinfo{journal}{Rev. Mod. Phys.} \textbf{\bibinfo{volume}{94}},
  \bibinfo{pages}{031003} (\bibinfo{year}{2022}),
  \eprint{doi:10.1103/RevModPhys.94.031003}.

\bibitem[{\citenamefont{Morgan et~al.}(2022)\citenamefont{Morgan, Pilania,
  Couet, Uberuaga, Sun, and Li}}]{Morgan2022review}
\bibinfo{author}{\bibfnamefont{D.}~\bibnamefont{Morgan}},
  \bibinfo{author}{\bibfnamefont{G.}~\bibnamefont{Pilania}},
  \bibinfo{author}{\bibfnamefont{A.}~\bibnamefont{Couet}},
  \bibinfo{author}{\bibfnamefont{B.~P.} \bibnamefont{Uberuaga}},
  \bibinfo{author}{\bibfnamefont{C.}~\bibnamefont{Sun}}, \bibnamefont{and}
  \bibinfo{author}{\bibfnamefont{J.}~\bibnamefont{Li}}, \bibinfo{journal}{Curr.
  Opin. Solid State Mater. Sci.} \textbf{\bibinfo{volume}{26}},
  \bibinfo{pages}{100975} (\bibinfo{year}{2022}),
  \eprint{doi:10.1016/j.cossms.2021.100975}.

\bibitem[{\citenamefont{Neudecker et~al.}(2020)\citenamefont{Neudecker,
  Grosskopf, Herman, Haeck, Grechanuk, Vander~Wiel, Rising, Kahler, Sly, and
  Talou}}]{neudecker2020enhancing}
\bibinfo{author}{\bibfnamefont{D.}~\bibnamefont{Neudecker}},
  \bibinfo{author}{\bibfnamefont{M.}~\bibnamefont{Grosskopf}},
  \bibinfo{author}{\bibfnamefont{M.}~\bibnamefont{Herman}},
  \bibinfo{author}{\bibfnamefont{W.}~\bibnamefont{Haeck}},
  \bibinfo{author}{\bibfnamefont{P.}~\bibnamefont{Grechanuk}},
  \bibinfo{author}{\bibfnamefont{S.}~\bibnamefont{Vander~Wiel}},
  \bibinfo{author}{\bibfnamefont{M.~E.} \bibnamefont{Rising}},
  \bibinfo{author}{\bibfnamefont{A.}~\bibnamefont{Kahler}},
  \bibinfo{author}{\bibfnamefont{N.}~\bibnamefont{Sly}}, \bibnamefont{and}
  \bibinfo{author}{\bibfnamefont{P.}~\bibnamefont{Talou}},
  \bibinfo{journal}{Nucl. Data Sheets} \textbf{\bibinfo{volume}{167}},
  \bibinfo{pages}{36} (\bibinfo{year}{2020}),
  \eprint{doi:10.1016/j.nds.2020.07.002}.

\bibitem[{\citenamefont{Cai et~al.}(2022)\citenamefont{Cai, Xu, {Di Lemma},
  Giglio, Benson, Murray, Adkins, Kane, Xian, Capriotti
  et~al.}}]{Cai2022MLnuclearfuels}
\bibinfo{author}{\bibfnamefont{L.}~\bibnamefont{Cai}},
  \bibinfo{author}{\bibfnamefont{F.}~\bibnamefont{Xu}},
  \bibinfo{author}{\bibfnamefont{F.~G.} \bibnamefont{{Di Lemma}}},
  \bibinfo{author}{\bibfnamefont{J.~J.} \bibnamefont{Giglio}},
  \bibinfo{author}{\bibfnamefont{M.~T.} \bibnamefont{Benson}},
  \bibinfo{author}{\bibfnamefont{D.~J.} \bibnamefont{Murray}},
  \bibinfo{author}{\bibfnamefont{C.~A.} \bibnamefont{Adkins}},
  \bibinfo{author}{\bibfnamefont{J.~J.} \bibnamefont{Kane}},
  \bibinfo{author}{\bibfnamefont{M.}~\bibnamefont{Xian}},
  \bibinfo{author}{\bibfnamefont{L.}~\bibnamefont{Capriotti}},
  \bibnamefont{et~al.}, \bibinfo{journal}{Materials Characterization}
  \textbf{\bibinfo{volume}{184}}, \bibinfo{pages}{111657}
  (\bibinfo{year}{2022}), ISSN \bibinfo{issn}{1044-5803},
  \eprint{doi:10.1016/j.matchar.2021.111657}.

\bibitem[{\citenamefont{Kautz et~al.}(2019)\citenamefont{Kautz, Hagen, Johns,
  and Burkes}}]{Kautz2019}
\bibinfo{author}{\bibfnamefont{E.~J.} \bibnamefont{Kautz}},
  \bibinfo{author}{\bibfnamefont{A.~R.} \bibnamefont{Hagen}},
  \bibinfo{author}{\bibfnamefont{J.~M.} \bibnamefont{Johns}}, \bibnamefont{and}
  \bibinfo{author}{\bibfnamefont{D.~E.} \bibnamefont{Burkes}},
  \bibinfo{journal}{Compl. Matls. Sci.} \textbf{\bibinfo{volume}{161}},
  \bibinfo{pages}{107} (\bibinfo{year}{2019}),
  \eprint{doi:10.1016/j.commatsci.2019.01.044}.

\bibitem[{\citenamefont{Xu et~al.}(2023)\citenamefont{Xu, Cai, Salvato,
  Capriotti, and Yao}}]{xu2023advanced}
\bibinfo{author}{\bibfnamefont{F.}~\bibnamefont{Xu}},
  \bibinfo{author}{\bibfnamefont{L.}~\bibnamefont{Cai}},
  \bibinfo{author}{\bibfnamefont{D.}~\bibnamefont{Salvato}},
  \bibinfo{author}{\bibfnamefont{L.}~\bibnamefont{Capriotti}},
  \bibnamefont{and} \bibinfo{author}{\bibfnamefont{T.}~\bibnamefont{Yao}},
  \bibinfo{journal}{Scientific Reports} \textbf{\bibinfo{volume}{13}},
  \bibinfo{pages}{10616} (\bibinfo{year}{2023}),
  \eprint{doi:10.1038/s41598-023-35619-1}.

\bibitem[{\citenamefont{Craven et~al.}(2023)\citenamefont{Craven, Chen, Cooper,
  Matthews, Rizk, Malone, Johnson, Gibson, and Andersson}}]{craven2023a}
\bibinfo{author}{\bibfnamefont{G.~T.} \bibnamefont{Craven}},
  \bibinfo{author}{\bibfnamefont{R.}~\bibnamefont{Chen}},
  \bibinfo{author}{\bibfnamefont{M.~W.} \bibnamefont{Cooper}},
  \bibinfo{author}{\bibfnamefont{C.}~\bibnamefont{Matthews}},
  \bibinfo{author}{\bibfnamefont{J.}~\bibnamefont{Rizk}},
  \bibinfo{author}{\bibfnamefont{W.}~\bibnamefont{Malone}},
  \bibinfo{author}{\bibfnamefont{L.}~\bibnamefont{Johnson}},
  \bibinfo{author}{\bibfnamefont{T.}~\bibnamefont{Gibson}}, \bibnamefont{and}
  \bibinfo{author}{\bibfnamefont{D.~A.} \bibnamefont{Andersson}},
  \bibinfo{journal}{Comput. Mater. Sci.} \textbf{\bibinfo{volume}{230}},
  \bibinfo{pages}{112442} (\bibinfo{year}{2023}),
  \eprint{doi:10.1016/j.commatsci.2023.112442}.

\bibitem[{\citenamefont{Cooper et~al.}(2023)\citenamefont{Cooper, Rizk,
  Matthews, Kocevski, Craven, Gibson, and Andersson}}]{craven2023b}
\bibinfo{author}{\bibfnamefont{M.~W.} \bibnamefont{Cooper}},
  \bibinfo{author}{\bibfnamefont{J.}~\bibnamefont{Rizk}},
  \bibinfo{author}{\bibfnamefont{C.}~\bibnamefont{Matthews}},
  \bibinfo{author}{\bibfnamefont{V.}~\bibnamefont{Kocevski}},
  \bibinfo{author}{\bibfnamefont{G.~T.} \bibnamefont{Craven}},
  \bibinfo{author}{\bibfnamefont{T.}~\bibnamefont{Gibson}}, \bibnamefont{and}
  \bibinfo{author}{\bibfnamefont{D.~A.} \bibnamefont{Andersson}},
  \bibinfo{journal}{J. Nucl. Mater.} \textbf{\bibinfo{volume}{587}},
  \bibinfo{pages}{154685} (\bibinfo{year}{2023}),
  \eprint{doi:10.1016/j.jnucmat.2023.154685}.

\bibitem[{\citenamefont{Craven and Wilson}(2023)}]{craven2023report}
\bibinfo{author}{\bibfnamefont{G.~T.} \bibnamefont{Craven}} \bibnamefont{and}
  \bibinfo{author}{\bibfnamefont{B.~M.} \bibnamefont{Wilson}},
  \bibinfo{type}{Tech. Rep.}, \bibinfo{institution}{Los Alamos National
  Laboratory} (\bibinfo{year}{2023}), \bibinfo{note}{{\it{G}eneralized kinetic
  model for defect evolution in irradiated materials}},
  \eprint{LA-UR-23-25179}.

\bibitem[{\citenamefont{Kröger and Vink}(1956)}]{KROGER1956307}
\bibinfo{author}{\bibfnamefont{F.}~\bibnamefont{Kröger}} \bibnamefont{and}
  \bibinfo{author}{\bibfnamefont{H.}~\bibnamefont{Vink}}
  (\bibinfo{publisher}{Academic Press}, \bibinfo{year}{1956}),
  vol.~\bibinfo{volume}{3} of \emph{\bibinfo{series}{Solid State Physics}}, pp.
  \bibinfo{pages}{307--435}.

\bibitem[{\citenamefont{Ronchi et~al.}(1978)\citenamefont{Ronchi, Ray, Thiele,
  and {van de Laar}}}]{Ronchi1978}
\bibinfo{author}{\bibfnamefont{C.}~\bibnamefont{Ronchi}},
  \bibinfo{author}{\bibfnamefont{I.}~\bibnamefont{Ray}},
  \bibinfo{author}{\bibfnamefont{H.}~\bibnamefont{Thiele}}, \bibnamefont{and}
  \bibinfo{author}{\bibfnamefont{J.}~\bibnamefont{{van de Laar}}},
  \bibinfo{journal}{J. Nucl. Mater.} \textbf{\bibinfo{volume}{74}},
  \bibinfo{pages}{193} (\bibinfo{year}{1978}), ISSN \bibinfo{issn}{0022-3115},
  \eprint{doi:10.1016/0022-3115(78)90359-8}.

\bibitem[{\citenamefont{Paszke et~al.}(2017)\citenamefont{Paszke, Gross,
  Chintala, Chanan, Yang, DeVito, Lin, Desmaison, Antiga, and Lerer}}]{Pytorch}
\bibinfo{author}{\bibfnamefont{A.}~\bibnamefont{Paszke}},
  \bibinfo{author}{\bibfnamefont{S.}~\bibnamefont{Gross}},
  \bibinfo{author}{\bibfnamefont{S.}~\bibnamefont{Chintala}},
  \bibinfo{author}{\bibfnamefont{G.}~\bibnamefont{Chanan}},
  \bibinfo{author}{\bibfnamefont{E.}~\bibnamefont{Yang}},
  \bibinfo{author}{\bibfnamefont{Z.}~\bibnamefont{DeVito}},
  \bibinfo{author}{\bibfnamefont{Z.}~\bibnamefont{Lin}},
  \bibinfo{author}{\bibfnamefont{A.}~\bibnamefont{Desmaison}},
  \bibinfo{author}{\bibfnamefont{L.}~\bibnamefont{Antiga}}, \bibnamefont{and}
  \bibinfo{author}{\bibfnamefont{A.}~\bibnamefont{Lerer}},
  \bibinfo{journal}{NIPS 2017 Workshop Autodif}  (\bibinfo{year}{2017}).

\bibitem[{\citenamefont{Zhou et~al.}(2021)\citenamefont{Zhou, Zhu, and
  Wu}}]{Zhou2021}
\bibinfo{author}{\bibfnamefont{X.-Y.} \bibnamefont{Zhou}},
  \bibinfo{author}{\bibfnamefont{J.-H.} \bibnamefont{Zhu}}, \bibnamefont{and}
  \bibinfo{author}{\bibfnamefont{H.-H.} \bibnamefont{Wu}},
  \bibinfo{journal}{Int. J. Hydrog. Energy} \textbf{\bibinfo{volume}{46}},
  \bibinfo{pages}{5842} (\bibinfo{year}{2021}), ISSN \bibinfo{issn}{0360-3199},
  \eprint{doi:10.1016/j.ijhydene.2020.11.131}.

\bibitem[{\citenamefont{Zhou et~al.}(2022)\citenamefont{Zhou, Zhu, Wu, Yang,
  Lookman, and Wu}}]{Zhou2022}
\bibinfo{author}{\bibfnamefont{X.-Y.} \bibnamefont{Zhou}},
  \bibinfo{author}{\bibfnamefont{J.-H.} \bibnamefont{Zhu}},
  \bibinfo{author}{\bibfnamefont{Y.}~\bibnamefont{Wu}},
  \bibinfo{author}{\bibfnamefont{X.-S.} \bibnamefont{Yang}},
  \bibinfo{author}{\bibfnamefont{T.}~\bibnamefont{Lookman}}, \bibnamefont{and}
  \bibinfo{author}{\bibfnamefont{H.-H.} \bibnamefont{Wu}},
  \bibinfo{journal}{Acta Mater.} \textbf{\bibinfo{volume}{224}},
  \bibinfo{pages}{117535} (\bibinfo{year}{2022}), ISSN
  \bibinfo{issn}{1359-6454}, \eprint{doi:10.1016/j.actamat.2021.117535}.

\end{thebibliography}

\end{document}